\documentstyle[12pt]{article}
\makeatletter 
\def\appendix{\par\clearpage 
  \setcounter{section}{0} 
  \setcounter{subsection}{0} 
  \@addtoreset{equation}{section} 
  \def\@sectname{Appendix~} 
  \def\theequation{\thesection.\arabic{equation}} 
  \def\thesection{\Alph{section}}} 
\makeatother 
 
\renewcommand{\theequation}{\thesection.\arabic{equation}} 
 
\begin{document} 
\begin{titlepage} 
\hskip 11cm \vbox{ 
\hbox{Budker INP/2000-63} 
\hbox{UNICAL-TH 00/3} 
\hbox{July 2000}} 
\vskip 0.3cm 
\centerline{\bf ONE-LOOP REGGEON-REGGEON-GLUON VERTEX}
\centerline{\bf AT ARBITRARY SPACE-TIME DIMENSION$^{~\ast}$} 
\vskip 0.8cm 
\centerline{V.S.~Fadin$^{a~\dagger}$, R.~Fiore$^{b~\ddagger}$, 
A.~Papa$^{b~\ddagger}$} 
\vskip .3cm 
\centerline{\sl $^a$ Budker Institute for Nuclear Physics,} 
\centerline{\sl Novosibirsk State University, 630090 Novosibirsk, Russia} 
\centerline{\sl $^b$ Dipartimento di Fisica, Universit\`a della Calabria,} 
\centerline{\sl Istituto Nazionale di Fisica Nucleare, Gruppo collegato di Cosenza,}
\centerline{\sl I-87036 Arcavacata di Rende, Cosenza, Italy} 
\vskip 1cm 
\begin{abstract} 
In order to check the compatibility of the gluon Reggeization in QCD with the 
$s$-channel unitarity, the one-loop correction to the Reggeon-Reggeon-gluon
vertex must be known at arbitrary space-time dimension $D$. We obtain this
correction from the gluon production amplitude in the multi-Regge kinematics
and present an explicit expression for it in terms of a few integrals over 
the transverse momenta of virtual particles. The one-gluon contribution to the
non-forward BFKL kernel at arbitrary $D$ is also obtained.
\end{abstract} 
\vfill 
\hrule 
\vskip.3cm 
\noindent 
$^{\ast}${\it Work supported in part by the Ministero italiano 
dell'Universit\`a  e della Ricerca Scientifica e Tecnologica, 
in part by INTAS and in part by the Russian Fund of Basic Researches.} 
\vfill 
$ \begin{array}{ll} 
^{\dagger}\mbox{{\it e-mail address:}} & 
 \mbox{FADIN ~@INP.NSK.SU}\\ 
\end{array} 
$ 
\vfill 
\vskip -1.5cm 
$ \begin{array}{ll} 
^{\ddagger}\mbox{{\it e-mail address:}} & 
 \mbox{FIORE, PAPA ~@FIS.UNICAL.IT}\\ 
\end{array} 
$ 
\vfill 
\vskip .1cm 
\vfill 
\end{titlepage} 
\eject 
 
\section{Introduction} 
\setcounter{equation}{0} 
 
Two years ago the kernel of the BFKL equation~\cite{BFKL} was obtained 
in the next-to-leading order (NLO)~\cite{FL98,CC} for the case of forward scattering,
i.e. for the momentum transfer $t=0$ and the singlet colour representation
in the $t$-channel. This long awaited event had led to the appearance of a number of
papers (see, for instance,~\cite{papers} and references therein) devoted to the 
problem of possible applications of the NLO result in the physics of semi-hard
processes.

An important way of development of the BFKL approach became the generalization
of the obtained results to the non-forward scattering~\cite{FF98}. For singlet
colour representation in the $t$-channel, the generalized approach can be used
directly for the description of a wide circle of physical processes. The
generalization for non-singlet colour states is also of great importance,
especially for the antisymmetric colour octet state of two Reggeized gluons in the 
$t$-channel. This case is especially important since the BFKL approach is
based on the gluon Reggeization, although the Reggeization is not proved
in the NLO. Therefore, at this order the Reggeization is a hypothesis which must
be carefully checked. This can be done using the ``bootstrap'' 
equations~\cite{FF98,BV} appearing from the requirement of the compatibility
of the gluon Reggeization with the $s$-channel unitarity. In the BFKL approach
the scattering amplitude for the high energy process $A+B \rightarrow A'+B'$ is
presented as the convolution of the impact factors $\Phi_{A'A}$ and
$\Phi_{B'B}$, which describe the $A \rightarrow A'$ and $B \rightarrow B'$
transitions in the particle-Reggeon scattering, and the Green's function
$G$ for the Reggeon-Reggeon scattering. The bootstrap conditions for
the impact factors are already checked both for the gluon and quark scattering,
for helicity conserving and helicity non-conserving 
amplitudes~\cite{FFKP00_G,FFKP00_Q}. We remind that the impact factors are infrared
finite~\cite{FM99} for colourless particles only. Considering scattering
amplitudes at the parton level, one needs to use an infrared regularization.
We use the dimensional regularization, which is commonly adopted. Remarkably, the
bootstrap conditions for the impact factors are satisfied at arbitrary space-time 
dimension $D=4+2 \epsilon$.

The kernel of the BFKL equation in the octet channel must also satisfy the bootstrap
condition (which is called ``first bootstrap condition''). This condition
was checked (also for arbitrary $D$) in the part concerning the
quark contribution to the kernel~\cite{FFP99}. Of course, the most important 
(and most complicated for verification) is the gluon part of the first 
bootstrap condition. The gluon part of the kernel in the NLO is expressed in terms
of the two-loop gluon trajectory, the one-loop effective vertex for the gluon 
production in the Reggeon-Reggeon collisions (RRG-vertex) and the two-gluon
production contribution. The last contribution was obtained recently~\cite{FG00}.
Since the two-loop gluon trajectory~\cite{FFK96a} and the one-loop 
RRG-vertex~\cite{FL93} are already known, it seems that the bootstrap condition
could be checked. But in the bootstrap equation~\cite{FF98}
\begin{equation}
\frac{g^2Nt}{2(2\pi)^{D-1}}\int \frac{d^{D-2}q_1}{\vec
q_1^{\:2}(\vec q_1-\vec q)^2}\int \frac{d^{D-2}q_2}{\vec
q_2^{\:2}(\vec q_2-\vec q)^2}{\cal K}^{(8)(1)}(\vec q_1,\vec q_2;\vec q)
=\omega^{(1)}(t)\,\omega^{(2)}(t)\;,
\label{1}
\end{equation}
where $g$ is the coupling constant, $N$ is the number of colour, $\vec q$ is
the transverse momentum transfer, $t=-\vec q^{\:2}$, ${\cal K}^{(8)(1)}$ is 
the one-loop contribution to the non-forward octet kernel, $\omega^{(1)}$
and $\omega^{(2)}$ are the one- and two-loop contributions to the gluon
trajectory, the integration over the transverse momenta $\vec q_{1,2}$
is singular. Therefore, in order to check at least the terms finite for
$\epsilon \rightarrow 0$ in $\omega^{(2)}$ in the R.H.S. of Eq.~(\ref{1}),
one needs to know the kernel in regions of singularities at arbitrary
$\epsilon$ (because, for example, the region of arbitrary small 
$\vec q_1$ does contribute, so that $(\vec q_1^{\:2})^\epsilon$ cannot 
be expanded in powers of $\epsilon$).
Unfortunately, the kernel with such accuracy is not known yet, although
for the gluon trajectory~\cite{FFK96a} and for the two-gluon contribution 
to the octet kernel~\cite{FG00} the integral representations for arbitrary 
$D$ are known. The problem is in the one-gluon contribution or, in other 
words, in the RRG-vertex. Let us note that, from now on, we will talk 
only about the gluon part, i.e. consider pure gluodynamics, since the 
quark part of the vertex is known at arbitrary $D$~\cite{FFQ94}. Instead, 
in the gluon part firstly only the
terms finite for $\epsilon\rightarrow0 $ were kept~\cite{FL93}, but in 
the process of calculation of the forward BFKL kernel, it was understood 
that in this kernel the RRG-vertex at small transverse momenta $\vec k$ 
of the produced gluon must be known at arbitrary $D$. After this, the 
vertex at small $\vec k$ was calculated~\cite{FFK96b} at 
arbitrary $D$. Later the results of~\cite{FL93,FFK96b} were 
obtained by another method in~\cite{DDS}. 
But for the verification of the
bootstrap equation~(\ref{1}) it must be known at arbitrary $D$ in a 
wider kinematical region. Therefore it became clear the necessity of  the 
knowledge of the RRG-vertex at arbitrary $D$, especially taking into 
account that it can be used not only for the check of the bootstrap, but, 
for example, in the Odderon problem in the NLO and so on.

In this paper we calculate the RRG-vertex in the NLO and its contribution to the 
non-forward BFKL kernel at arbitrary $D$. In the next Section we calculate
the amplitude of the gluon production in the multi-Regge kinematics for gluon-gluon
collisions. The RRG-vertices are defined and calculated in Section~3 and the
one-gluon contribution to the BFKL kernel in Section~4.

\section{The gluon production amplitude} 
\setcounter{equation}{0} 

The RRG-vertex can be obtained from the amplitudes of the gluon production
in the multi-Regge kinematics (in other words, in the central kinematical 
region) at collisions of any pair of particles. We will
consider the gluon-gluon collisions and use intermediate results obtained
in~\cite{FL93} for this process, which are valid at arbitrary $D$. We will 
use the denotations $p_A$ and $p_B$ ($p_{A'}$ and $p_{B'}$) for the 
momenta of the incoming (outgoing) gluons, $k$ and $e(k)$ for momentum 
and polarization vector of the produced gluon; 
$q_1= p_A - p_{A'}$ and $q_2= p_{B'} - p_B$ are the momentum transfers, so that 
$k=q_1-q_2$. All the denotations are the same as used in~\cite{FL93}, except 
that for the momentum of the produced gluon, which was called $p_D$ there. The
kinematics is defined by the relations
\begin{equation}
s \gg s_1\sim s_2 \gg |t_1| \sim |t_2|\;,
\label{2}
\end{equation}
where
\begin{equation}
s=(p_A+p_B)^2\;, \;\;\;\;\; s_1=(p_{A'}+k)^2\;, \;\;\;\;\; 
s_2=(p_{B'}+k)^2\;, 
\;\;\;\;\; t_{1,2}=q_{1,2}^2\;.
\label{3}
\end{equation}
In terms of the parameters of the Sudakov decomposition
\begin{equation}
k = \beta \, p_A + \alpha \, p_B + k_\perp \;, \;\;\;\;\;\;\;\;\;\;
q_i = \beta_i \, p_A + \alpha_i \, p_B + {q_i}_{\perp} \;, 
\label{4}
\end{equation}
the relations~(\ref{2}) give
\[
1 \gg \beta \approx \beta_1 \gg -\alpha_1 \simeq \frac{\vec q_1^{\:2}}{s} \;,
\;\;\;\;\;\;\;\;\;\;
1 \gg \alpha \approx -\alpha_2 \gg \beta_2 \simeq\frac{\vec q_2^{\:2}}{s} \;,
\]
\begin{equation}
s_1 \approx s \, \alpha \;, \;\;\;\;\; s_2 \approx s \, \beta \;, \;\;\;\;\;
\vec k^2 = - k_\perp^2 \approx \frac{s_1 s_2}{s}\;.
\label{5}
\end{equation}
Here and below the vector sign is used for the components of the momenta transverse
to the plane of the momenta of the initial particles $p_A$ and $p_B$. 

Since we are interested in the RRG-vertex, we will consider the amplitudes with
conservation of the helicities of the scattered gluons (helicity non-conservation
in the NLO is related with the gluon-gluon-Reggeon vertices only, which are
already known at arbitrary $D$~\cite{FL93}). This amplitude can be presented
in the form (cf.~\cite{FL93})
\begin{equation}
A_{2\rightarrow 3} = 2s \, g^3\, T_{A'A}^{c_1} \, \frac{1}{t_1} \, T_{c_2c_1}^{d} 
\, \frac{1}{t_2}\,T_{B'B}^{c_2} \, e^*_\mu(k) \, {\cal A}^\mu(q_2,q_1)\;,
\label{6}
\end{equation} 
where $T_{bc}^a$ are matrix elements of the colour group generator 
in the adjoint representation and the amplitude
${\cal A}^\mu$ in the Born approximation is equal to $C^\mu(q_2,q_1)$: 
\begin{equation}
{\cal A}_{\mathrm Born}^\mu = C^\mu(q_2,q_1) = -{q_1}_{\perp}^\mu -{q_2}_{\perp}^\mu
+\frac{p_A^\mu}{s_1}\,(\vec k^2 - 2\vec q_1^{\:2})
-\frac{p_B^\mu}{s_2}\,(\vec k^2 - 2\vec q_2^{\:2}) \; .
\label{7}
\end{equation}

It was shown in~\cite{FL93} that at one-loop order the amplitude can be presented
as the sum of three contributions, which come from integrations over the momenta of 
the virtual gluons in three different kinematical regions: the region of
fragmentation of the particle $A$, the central region and the region of
fragmentation of the particle $B$. Correspondingly, we represent ${\cal A}^\mu$
as 
\begin{equation}
{\cal A}^\mu = {\cal A}_{\mathrm Born}^\mu + {\cal A}_A^\mu 
+ {\cal A}_{\mathrm as}^\mu + {\cal A}_B^\mu \; ,
\label{8}
\end{equation}
where for the contribution of the fragmentation of the particle $A$, we obtain
from Eq.~(63) of~\cite{FL93}
\begin{equation}
{\cal A}_A^\mu = C^\mu(q_2,q_1)\,\overline g^2 \, (\vec q_1^{\:2})^\epsilon \,
\frac{\Gamma^2(1+\epsilon)}{\Gamma(4+2\epsilon)} \, \frac{(-2)}{\epsilon^2} \,
\biggl(3(1+\epsilon)^2+\epsilon^2\biggr) \;,
\label{9}
\end{equation}
where $\overline g^2 = g^2 N \Gamma(1-\epsilon)/(4 \pi)^{2+\epsilon}$. In the 
derivation of~(\ref{9}) from Eq.~(63) of~\cite{FL93}, we have used the 
relations
\begin{equation}
e_{A\lambda}^\alpha \, e_{A'\lambda}^{*\alpha'} \, 
\left(\delta_{\perp\alpha\alpha'}-(D-2) 
\frac{{q_1}_{\perp\alpha} \, {q_1}_{\perp\alpha'}}{{q_1}_{\perp}^2}\right) = 0\;, 
\;\;\;\;\;\;\;\;\;\;
e_{A\lambda}^\alpha \, e_{A'\lambda}^{*\alpha'}\, 
\delta_{\perp\alpha\alpha'} = -1\, ,
\label{10}
\end{equation}
for the polarization vectors $e_{A\lambda}$
and $e_{A'\lambda}$ of the particles $A$ and $A'$ with the definite 
helicity $\lambda$ in the $D$-dimensional space-time. The analogous relations exist
for the particle $B$. Evidently, the contribution from the fragmentation
region of the particle $B$ can be obtained from~(\ref{9}) by the substitution
$\vec q_1^{\:2} \rightarrow \vec q_2^{\:2}$.

The contribution of the central region is given by Eqs.~(77),~(80) and~(81) 
of~\cite{FL93}. At this step in~\cite{FL93} was not yet made an expansion
in $\epsilon$. Let us pay attention that in these formulae the 
anti-symmetrization with respect to  the substitution 
$p_B \longleftrightarrow -p_{B'}$ must be done. The formulae~(80) 
and~(81) of~\cite{FL93} can be greatly simplified 
and the contribution ${\cal A}_{\mathrm as}^\mu$, with accuracy up to terms
proportional to $k^\mu$ and therefore not contributing to~(\ref{8}), 
can be presented as
\[
{\cal A}_{\mathrm as}^\mu = -\frac{2 g^2 N}{(2 \pi)^D} \, \widehat {\cal S}
\left\{\biggl(C^\mu(q_2,q_1)-2 \vec k^2 {\cal P}^\mu\biggr) \,
\left[s_1 t_1 I_{4A}-\frac{4(D-2)}{t_1} \frac{p_A^\rho p_B^\sigma}{s}
I_2^{\rho\sigma}(q_1) - 2 t_1 \frac{p_B^\rho}{s}I_{3A}^\rho \right.\right.
\]
\[
\left.
+ 4 \vec k^2 \frac{p_A^\rho}{s_1} I_3^\rho -2 t_1 I_{3A} + \vec k^2 I_3 - 2I_2(q_1)
\right] + 4 {\cal P}^\mu \left[\frac{s_1 s_2}{4} t_1 t_2 I_5 
+ \frac{s_1 t_1}{2}(\vec k^2-t_2) I_{4A} - 2 \vec k^2 t_1 \frac{p_A^\rho}{s_1}
I_3^\rho \right.
\]
\[
\left.
- \vec k^2 t_2 I_3 - \vec k^2 \biggl(2 I_2(q_1) + t_1 I_{3A}\biggr) 
-t_2 I_2(q_1)\right] -2 \left(g^{\mu\rho}
-2 \frac{p_B^\mu k^\rho}{s_2}\right)\left[\frac{t_1 s_2}{s} I_{3A}^\rho \right.
\]
\[
\left.
+ 2 (D-2) \frac{\vec k^2}{t_1} \frac{p_A^\sigma}{s_1} 
I_2^{\rho\sigma}(q_1)\right] - 4(D-2) \frac{p_A^\rho p_B^\sigma}{s} 
\left[I_3^{\mu\rho\sigma}+I_3^{\mu\rho}k^\sigma - 2 \frac{p_A^\mu}{s_1} 
I_2^{\rho\sigma}(q_1)\right]
\]
\begin{equation}
\left.
-s t_1 t_2 \left[2\frac{p_B^\mu}{s_2}I_4^{(1)} + I_5^\mu\right]
+2 (\vec k^2+t_1+t_2) \left[2\frac{p_A^\mu}{s_1}I_2(q_1) - I_3^\mu\right]
\right\}\;,
\label{11}
\end{equation}
where 
\begin{equation}
{\cal P}^\mu = \frac{p_A^\mu}{s_1}-\frac{p_B^\mu}{s_2}
\label{13}
\end{equation}
and $\widehat{\cal S}$ is the operator of symmetrization with respect 
to each 
of the substitutions $p_A \longleftrightarrow - p_A$ and 
$p_B \longleftrightarrow - p_B$
and anti-symmetrization with respect to the substitution 
\begin{equation}
p_A \longleftrightarrow p_B \;, \;\;\;\;\;\;\;\;\;\; q_1 \longleftrightarrow 
- q_2\;.
\label{12}
\end{equation}
The $I$'s in Eq.~(\ref{11}) are the integrals introduced in~\cite{FL93}:
\[
I_2^{\mu_1 \cdots \mu_n}(q) = \frac{1}{i} \int \, d^D p \, \frac{p^{\mu_1}
\cdots p^{\mu_n}}{(p^2+i\varepsilon)[(p+q)^2+i\varepsilon]} \;,
\]
\[
I_3^{\mu_1 \cdots \mu_n} = \frac{1}{i} \int \, d^D p \, \frac{p^{\mu_1}
\cdots p^{\mu_n}}{(p^2+i\varepsilon)[(p+q_1)^2+i\varepsilon] 
[(p+k)^2+i\varepsilon]} \;,
\]
\[
I_{3A}^{\mu_1 \cdots \mu_n} = \frac{1}{i} \int \, d^D p \, \frac{p^{\mu_1}
\cdots p^{\mu_n}}{(p^2+i\varepsilon)[(p+q_1)^2+i\varepsilon] 
[(p+p_A)^2+i\varepsilon]} \;,
\]
\begin{equation}
I_4^{(1)} = \frac{1}{i} \int \, d^D p \, \frac{1}
{(p^2+i\varepsilon)[(p+q_1)^2+i\varepsilon][(p+p_A)^2+i\varepsilon]
[(p-p_B)^2+i\varepsilon]} \;,
\label{14}
\end{equation}
\[
I_{4A} = \frac{1}{i} \int \, d^D p \, \frac{1}
{(p^2+i\varepsilon)[(p+q_1)^2+i\varepsilon][(p+q_2)^2+i\varepsilon]
[(p+p_A)^2+i\varepsilon]} \;,
\]
\[
I_5 = \frac{1}{i} \int \, d^D p \, \frac{1}
{(p^2+i\varepsilon)[(p+q_1)^2+i\varepsilon][(p+q_2)^2+i\varepsilon]
[(p+p_A)^2+i\varepsilon][(p-p_B)^2+i\varepsilon]} \;,
\]
\[
I_5^\mu = \frac{1}{i} \int \, d^D p \, \frac{(p+q_1)^\mu}
{(p^2+i\varepsilon)[(p+q_1)^2+i\varepsilon][(p+q_2)^2+i\varepsilon]
[(p+p_A)^2+i\varepsilon][(p-p_B)^2+i\varepsilon]} \;.
\]
In the representation~(\ref{11}) the transversality to the momenta $k^\mu$
(which guarantees the gauge invariance of the amplitude) of the first three terms
with square brackets is evident, while the transversality of the last three
terms can be easily checked with account of the anti-symmetrization~(\ref{12}).

The integrals with two or three denominators ($I_2^{\mu_1 \cdots \mu_n}$,
$I_3^{\mu_1 \cdots \mu_n}$, $I_{3A}^{\mu_1 \cdots \mu_n}$) were calculated
in~\cite{FL93} at arbitrary $D$, whereas for the other integrals the expansion in
$\epsilon$ was used. It occurs that, in the multi-Regge kinematics~(\ref{2}), 
the integral $I_4^{(1)}$ can also be calculated at arbitrary $D$, as it is 
shown in Appendix~A. Unfortunately, the integrals $I_{4A}$, $I_5$ and
$I_5^\mu$ cannot be expressed in terms of elementary functions at arbitrary $D$.
It is possible, however, to perform the integration over the longitudinal variables
in the Sudakov decomposition for $p$ and to express them in terms of the same 
$(D-2)$-dimensional integrals over $\vec p$, which appear in the two-gluon
contribution to the kernel~\cite{FG00}. This is done in Appendix~A, where,
for completeness, the integrals with two and three denominators are also given.

The integrals which cannot be expressed through elementary functions, and will be 
used below, are the following:
\[
{\cal I}_{4A} \!\!=\!\! \int_0^1 \!\!\!\frac{dx}{1-x}\! \int\!\!\! 
\frac{d^{2+2\epsilon}k_1}
{\pi^{1+\epsilon} \Gamma(1-\epsilon)} \! \left[\frac{x}{[(1-x) \vec k_1^2+
x (\vec k_1+\vec q_2)^2](\vec k_1-x\vec k)^2} 
-\frac{1}{(\vec k_1+\vec q_2)^2 (\vec k_1-\vec k)^2}\right]\!,
\]
\[
{\cal I}_5 = \int_0^1 \frac{dx}{1-x} \int \frac{d^{2+2\epsilon}k_1}
{\pi^{1+\epsilon} \Gamma(1-\epsilon)} \frac{1}{\vec k_1^2 [(1-x) \vec k_1^2+
x (\vec k_1-\vec q_1)^2]} \left(\frac{x^2}{(\vec k_1-x\vec k)^2}-\frac{1}
{(\vec k_1-\vec k)^2}\right) \;,
\]
\[
{\cal L}_3 = \int \frac{d^{2+2\epsilon}k_1}
{\pi^{1+\epsilon} \Gamma(1-\epsilon)} \frac{1}{\vec k_1^2 (\vec k_1-\vec q_1)^2
(\vec k_1-\vec q_2)^2} \ln\left(\frac{(\vec k_1-\vec q_1)^2 (\vec k_1-\vec q_2)^2}
{\vec k^2 \, \vec k_1^2}\right)\;,
\]
\begin{equation}
{\cal I}_3 = \int \frac{d^{2+2\epsilon}k_1}
{\pi^{1+\epsilon} \Gamma(1-\epsilon)} \frac{1}{\vec k_1^2 (\vec k_1-\vec q_1)^2
(\vec k_1-\vec q_2)^2} \;,
\label{15}
\end{equation}
as well as ${\cal I}_{4B}$, which is obtained from ${\cal I}_{4A}$ by the
substitution $q_1 \longleftrightarrow -q_2$, and ${\cal I}_5^\mu$, ${\cal L}_3^\mu$
and ${\cal I}_3^\mu$, the first of which differs from the integral ${\cal I}_5$
by the factor $(k_1-q_1)_\perp^\mu$ and the other two from ${\cal L}_3$
and ${\cal I}_3$, respectively, by the factor ${k_1}_{\perp}^\mu$, in the 
corresponding integrands.

Using the results of the Appendix~A, one gets 
\begin{equation}
{\cal A}_{\mathrm as}^\mu = \overline g ^2 \left\{ - 2 t_1 t_2 {\cal F}_5^\mu
+ r_{\mathrm as} C^\mu(q_2,q_1) + 2 t_1 r_A \frac{p_A^\mu}{s_1}
- 2 t_2 r_B \frac{p_B^\mu}{s_2}\right\}\;,
\label{16}
\end{equation}
where
\[
{\cal F}_5^\mu = {\cal I}_5^\mu  + {\cal L}_3^\mu + \frac{1}{2} 
\ln\left(\frac{s (-s) (\vec k^2)^2}{s_1 (-s_1) s_2 (-s_2)}\right) {\cal I}_3^\mu\;,
\]
\[
r_{\mathrm as} = \left\{ \frac{t_1 t_2}{2} {\cal F}_5 + t_2 {\cal I}_{4B}
+ \frac{\Gamma^2(\epsilon)}{\Gamma(2\epsilon)} (\vec q_1^{\:2})^\epsilon
\left[ -\frac{1}{2}\ln\left(\frac{s_1 (-s_1)}{t_1^2}\right) 
+2 \psi(\epsilon) - \psi(2\epsilon) - \psi(1-\epsilon)
\right. \right.
\]
\[
+\frac{1}{2 \epsilon (1+2\epsilon) (3+2\epsilon)}
\left(\frac{t_1(3+14\epsilon+8\epsilon^2)-t_2(3+3\epsilon+\epsilon^2)}{t_1-t_2}
\right.
\]
\[
\left.\left.\left.
+ \frac{\vec k^2 t_1\, \epsilon}{(t_1-t_2)^3} \biggl((2+\epsilon)t_2-\epsilon \, t_1
\biggr)\right)\right]\right\} 
+ \biggl\{ A \longleftrightarrow B\biggr\}\;,
\]
\[
{\cal F}_5 = {\cal I}_5 - {\cal L}_3 - \frac{1}{2} 
\ln\left(\frac{s (-s) (\vec k^2)^2}{s_1 (-s_1) s_2 (-s_2)}\right) {\cal I}_3\;,
\]
\[
r_A = -t_2 (t_1 {\cal F}_5 + {\cal I}_{4B}) + \frac{\Gamma^2(\epsilon)}
{\Gamma(2\epsilon)} (\vec q_2^{\:2})^\epsilon
\left[ \frac{1}{2}\ln\left(\frac{s_1 (-s_1)s_2 (-s_2)}{s(-s)t_2^2}\right) 
-\psi(1) - \psi(\epsilon) \right.
\]
\[
\biggl.
+ \psi(1-\epsilon)+ \psi(2\epsilon)\biggr]
+ \frac{\Gamma^2(\epsilon)}{2 t_1 (1+2\epsilon)\Gamma(2\epsilon) (3+2\epsilon)}
\left\{\left((\vec q_1^{\:2})^{\epsilon+1}\left[\frac{t_2}{t_1-t_2}(11+7\epsilon)
\right. \right.\right.
\]
\[
\left.\left.\left.
+ \frac{\vec k^2}{(t_1-t_2)^3}\biggl(t_2(t_1+t_2)-\epsilon \, t_1(t_1-t_2)\biggr)
+\frac{(\vec k^2)^2}{(t_1-t_2)^3} \biggl((2+\epsilon) t_2 - \epsilon \, t_1\biggr)
\right]\right) + \left(A \longleftrightarrow B\right)\right\}\;,
\]
\begin{equation}
r_B = r_A (A \longleftrightarrow B) \;.
\label{17}
\end{equation}
Here and below $A \longleftrightarrow B$ means the substitution~(\ref{12}).
The amplitude ${\cal A}^\mu$ must be anti-symmetric under this substitution. The 
amplitude~(\ref{16}) has this property since $C^\mu(q_2,q_1)$ (Eq.~(\ref{7}))
and ${\cal F}_5^\mu$ (Eqs.~(\ref{17})) change their sign under this substitution.

Note that in all physical regions we have
$s_1 s_2/s = \vec k^2$, so that 
\begin{equation}
\ln\left(\frac{s(-s)(\vec k^2)^2}{(s_1(-s_1) s_2 (-s_2))}\right) = i \,\pi\;.
\label{ip}
\end{equation}
Below we will work in the physical region of the $s$-channel and use this relation.

The total amplitude ${\cal A}^\mu$ is defined by~(\ref{8}),~(\ref{9}) 
and~(\ref{16}) and can be represented as 
\begin{equation}
{\cal A}^\mu =C^\mu(q_2,q_1)\left(1+\overline g ^2 r\right)+
 \overline g ^2 \left\{ - 2 t_1 t_2 {\cal F}_5^\mu
 + 2 t_1 r_A \frac{p_A^\mu}{s_1}
- 2 t_2 r_B \frac{p_B^\mu}{s_2}\right\}\;,
\label{16a}
\end{equation}
where
\begin{equation}
r =  r_{\mathrm as} -\frac{2}{\epsilon^2}
\frac{\Gamma^2(1+\epsilon)}{\Gamma(4+2\epsilon)} 
\biggl(3(1+\epsilon)^2+\epsilon^2\biggr)
\biggl((\vec q_1^{\:2})^\epsilon+(\vec q_2^{\:2})^\epsilon\biggr)
\label{19}
\end{equation}
and $r_{\mathrm as}$, $r_A$ and $r_B$ are defined in (\ref{17}). 
The gauge invariance of the amplitude~(\ref{6}) follows from the properties of the 
amplitudes~(\ref{9}) and~(\ref{16}):
\begin{equation}
k_\mu {\cal A}_A^\mu = k_\mu {\cal A}_B^\mu = k_\mu {\cal A}_{\mathrm as}^\mu=0\;.
\label{20}
\end{equation}
The first two of these relations follow evidently from the transversality
of $C^\mu(q_2,q_1)$ (Eq.~(\ref{7})), whereas the last of them is not so trivial and 
is fulfilled due to the equality:
\[
2 \, k_\mu \, {\cal F}_5^\mu \equiv f_{_-} = \left[-t_1 {\cal F}_5 - {\cal I}_{4B}
- \frac{\Gamma^2(\epsilon)}{\Gamma(2\epsilon)}(\vec q_2^{\: 2})^{\epsilon-1}
\left(\frac{1}{2} \ln\left(\frac{s_1 (-s_1) s_2(-s_2)}{s(-s) t_2^2}\right)
\right.\right.
\]
\begin{equation}
\biggl. \biggl.
+ \psi(1-\epsilon) - \psi(\epsilon) + \psi(2 \epsilon) - \psi(1) \biggr)\biggr]
- \biggl[ A \longleftrightarrow B \biggr] \;, 
\label{21}
\end{equation}
which is derived in the Appendix~B. The amplitude can be written in an explicit 
gauge invariant form if we use an analogous relation, also obtained in the 
Appendix~B:
\[
2 \, (q_1+q_2)_\mu \, {\cal F}_5^\mu \equiv f_{_+} = \left\{-t_1 {\cal F}_5 
-{\cal I}_{4B}-\frac{\Gamma^2(\epsilon)}{\Gamma(2\epsilon)}
\left[(\vec q_2^{\: 2})^{\epsilon-1}
\left(\frac{1}{2} \ln\left(\frac{s_1 (-s_1) s_2(-s_2)}{s(-s) t_2^2}\right)
\right.\right.\right.
\]
\[
\biggl. 
+ \psi(1-\epsilon) - \psi(\epsilon) + \psi(2 \epsilon) - \psi(1) \biggr)
\]
\begin{equation}
\left.\left.
-(\vec k^2)^{\epsilon-1}
\left(\frac{1}{2} \ln\left(\frac{s_1 (-s_1) s_2(-s_2)}{s(-s) (\vec k^2)^2}
\right) + \psi(1-\epsilon) - \psi(\epsilon)\right)\right]\right\}
+ \biggl\{ A \longleftrightarrow B \biggr\} \;.
\label{22}
\end{equation}
Since the general form for ${\cal F}_5^\mu$ is
\begin{equation}
{\cal F}_5^\mu = r_{_-} \, k_\perp^\mu + r_{_+} \, (q_1+q_2)_\perp^\mu \;,
\label{23}
\end{equation}
we can express $r_{_-}$ and $r_{_+}$ in terms of $f_{_-}$ and $f_{_+}$:
\begin{equation}
r_{_+} = \frac{f_{_-} (\vec q_1^{\:2}-\vec q_2^{\:2})-f_{_+} \vec k^2}
{8 (\vec q_1^{\:2} \vec q_2^{\:2} - (\vec q_1 \vec q_2)^2)} \;,
\;\;\;\;\;\;\;\;\;\;
r_{_-} = \frac{f_{_+} (\vec q_1^{\:2}-\vec q_2^{\:2})-f_{_-} (\vec q_1+\vec q_2)^2}
{8 (\vec q_1^{\:2} \vec q_2^{\:2} - (\vec q_1 \vec q_2)^2)} \;.
\label{24}
\end{equation}
Rewriting then~(\ref{23}) as
\begin{equation}
{\cal F}_5^\mu = r_{_-} \,k^\mu - r_{_+}\, \left[C^\mu(q_2,q_1)+ 
2\vec q_1\vec q_2 \left(\frac{p_A^\mu}{s_1}-\frac{p_B^\mu}{s_2}\right)\right]
+\frac{f_{_-}}{2}\left(\frac{p_A^\mu}{s_1}+\frac{p_B^\mu}{s_2}\right)
\label{25}
\end{equation}
and using this equality in the expression~(\ref{16}) for ${\cal A}_{\mathrm as}^\mu$,
we obtain the explicitly gauge invariant form for the amplitude ${\cal A}^\mu$:
\begin{equation}
{\cal A}^\mu = C^\mu(q_2,q_1) (1+\overline g^2 r_C) + {\cal P}^\mu \overline g^2
\, 2 t_1 t_2 \, r_{\cal P}\;,
\label{26}
\end{equation}
where the terms proportional to $k^\mu$ were omitted and 
\[
r_C = \left\{t_1 t_2 \left(r_{_+}+\frac{{\cal F}_5}{2}\right)+t_2{\cal I}_{4B} 
+ \frac{\Gamma^2(\epsilon)}{\Gamma(2\epsilon)} (\vec q_1^{\:2})^\epsilon
\left[ -\frac{1}{2}\ln\left(\frac{s_1 (-s_1)}{t_1^2}\right)
+2 \psi(\epsilon) - \psi(2\epsilon)
\right.\right.
\] 
\[
- \psi(1-\epsilon)+\frac{1}{2\epsilon(1+2\epsilon)(3+2\epsilon)}\left(-3(1+\epsilon)
-\frac{\epsilon^2}{1+\epsilon} + \frac{t_1(3+14\epsilon+8\epsilon^2)
-t_2(3+3\epsilon+\epsilon^2)}{t_1-t_2} \right. 
\]
\[
\left.\left.
+ \frac{\vec k^2 \, t_1 \, \epsilon}{(t_1-t_2)^3}
\biggl((2+\epsilon)t_2-\epsilon\,t_1\biggr)\right]\right\}
+ \biggl\{A \longleftrightarrow B \biggr\}\;,
\]
\[
r_{\cal P} = \left\{(\vec q_1 \vec q_2) r_{_+} - \frac{t_1+t_2}{4}{\cal F}_5
-\frac{{\cal I}_{4B}}{2} 
+ \frac{1}{2}\frac{\Gamma^2(\epsilon)}{\Gamma(2\epsilon)} 
(\vec q_1^{\:2})^{\epsilon-1}
\left( -\frac{1}{2}\ln\left(\frac{s_1(-s_1)s_2(-s_2)}{s(-s)t_1^2}\right)
+ \psi(1)
\right.\right.
\]
\[
+\psi(\epsilon) - \psi(1-\epsilon) - \psi(2\epsilon)
+\frac{t_1}{t_2\,(1+2\epsilon)(3+2\epsilon)}
\left[ \frac{t_2}{t_1-t_2} (11+7\epsilon) \right.
\]
\begin{equation}
\left.\left.\left.
+ \frac{\vec k^2}{(t_1-t_2)^3}\biggl(t_2(t_1+t_2)-\epsilon\,t_1(t_1-t_2)\biggr)
+ \frac{(\vec k^2)^2}{(t_1-t_2)^3}\biggl((2+\epsilon)t_2-\epsilon\,t_1\biggr)
\right]\right)\right\} + \biggl\{ A \longleftrightarrow B \biggr\}\;.
\label{27}
\end{equation}
The functions $r_{_+}$, ${\cal F}_5$ and ${\cal I}_{4B}$ entering
Eq.~(\ref{27}) are given by~(\ref{21})-(\ref{24}),~(\ref{17}) and~(\ref{15}),
respectively.

\section{The Reggeon-Reggeon-gluon vertices} 
\setcounter{equation}{0} 

Since the production amplitude must not have simultaneously discontinuities 
in the overlapping channels $s_1$ and $s_2$, it cannot be a simple 
generalization of the Regge form for the elastic amplitude. Instead, it has
the form~\cite{FL93,bartels80}
\[
8\, g^3 \, {\cal A}^\mu = \Gamma(t_1;\vec k^2) \Gamma(t_2;\vec k^2)
\left\{
\left[\left(\frac{s_1}{\vec k^2}\right)^{\omega_1-\omega_2}
+\left(\frac{-s_1}{\vec k^2}\right)^{\omega_1-\omega_2}\right]
\left[\left(\frac{s}{\vec k^2}\right)^{\omega_2}
+\left(\frac{-s}{\vec k^2}\right)^{\omega_2}\right]\, R^\mu \right.
\]
\begin{equation}
\left.
+ \left[\left(\frac{s_2}{\vec k^2}\right)^{\omega_2-\omega_1}
+\left(\frac{-s_2}{\vec k^2}\right)^{\omega_2-\omega_1}\right]
\left[\left(\frac{s}{\vec k^2}\right)^{\omega_1}
+\left(\frac{-s}{\vec k^2}\right)^{\omega_1}\right]\, L^\mu 
\right\}\;,
\label{28}
\end{equation}
where $\omega_i = \omega(t_i)$ and we have chosen $\vec k^2$ as the scale 
of energy, since with this choice the one-gluon contribution to the 
BFKL kernel is expressed through the RRG-vertex in the simplest 
way~\cite{rio98}. $\Gamma(t_i;\vec k^2)$ are the helicity conserving 
gluon-gluon-Reggeon vertices; they depend on $\vec k^2$ as on the
energy scale~\cite{rio98}. $R^\mu$ and $L^\mu$ are the right and left 
RRG-vertices, depending on $\vec q_1$ and $\vec q_2$. They are real 
in all physical channels, as well as $\Gamma(t_i; \vec k^2)$. 

In the one-loop approximation we have 
\begin{equation}
\omega(t) = \omega^{(1)}(t) = - \overline g^2 \frac{\Gamma^2(\epsilon)}
{\Gamma(2\epsilon)} (\vec q^{\:2})^\epsilon\;,
\label{29}
\end{equation}
\begin{equation}
\Gamma(t;\vec k^2) =  g \biggl(1+\Gamma^{(1)}(t;\vec k^2) \biggr)\;,
\label{30}
\end{equation}
where
\[
\Gamma^{(1)}(t;\vec k^2) = \overline g^2 \frac{\Gamma^2(\epsilon)}
{\Gamma(2\epsilon)} (\vec q^{\:2})^\epsilon \left[\psi(\epsilon)
- \frac{1}{2} \psi(1) - \frac{1}{2} \psi(1-\epsilon) \right.
\]
\begin{equation}
\left.
+\frac{9 (1+\epsilon)^2+2}{4(1+\epsilon) (1+2\epsilon) (3+2\epsilon)}
-\frac{1}{2} \ln\left(\frac{\vec k^2}{\vec q^{\:2}}\right)\right] \;.
\label{31}
\end{equation}
In the same approximation we obtain from~(\ref{28})
\[
2 g \left\{{\cal A}^\mu - C^\mu(q_2,q_1)\left[\Gamma^{(1)}(t_1;\vec k^2)
+\Gamma^{(1)}(t_2;\vec k^2)
+\frac{\omega_1}{2}\ln\left(\frac{s_1(-s_1)}{(\vec k^2)^2}\right)
+\frac{\omega_2}{2}\ln\left(\frac{s_2(-s_2)}{(\vec k^2)^2}\right) \right.\right.
\]
\begin{equation}
\left.\left.
+\frac{\omega_1+\omega_2}{4}\ln\left(\frac{s(-s)(\vec k^2)^2}{s_1(-s_1)s_2(-s_2)}
\right)\right]\right\}
= R^\mu + L^\mu + (R^\mu - L^\mu) \frac{\omega_1-\omega_2}{4} 
\ln\left(\frac{s_1(-s_1)s_2(-s_2)}{s(-s)(\vec k^2)^2}\right)\;,
\label{32}
\end{equation}
where now $\omega_i\equiv \omega^{(1)}(t_i)$.
Since in the physical region of the $s$-channel the relation~(\ref{ip}) holds, 
the combinations $R^\mu + L^\mu$ and $R^\mu - L^\mu$ are determined by the real
and imaginary parts of the L.H.S. of Eq.~(\ref{32}) in this region. 
Using~(\ref{16a}) and comparing the imaginary parts in~(\ref{32}), we 
obtain: 
\[
R^\mu - L^\mu =  \frac{2g \overline g^2 }{\omega_1-\omega_2}
\left\{\left[2t_1t_2{\cal I}_{3}^{\mu}+ \left(C^\mu(q_2,q_1)
-4t_1\frac{p_A^{\mu}}{s_1}\right)
\right.\right.
\]
\begin{equation}
\left.\left.\times \left(t_1t_2{\cal I}_3-\frac{\Gamma^2(\epsilon)}
{\Gamma(2\epsilon)}(\vec q_2^{\:2})^{\epsilon}\right)\right]
-\left[A\longleftrightarrow B\right]\right\}\;.
\label{33}
\end{equation}
We can express ${\cal I}_{3}^{\mu}$ in term of ${\cal I}_{3}$ 
by comparing the imaginary parts of~(\ref{25}). In this way we obtain 
from~(\ref{33}) the explicitly gauge invariant expression: 
\[
R^\mu - L^\mu =  \frac{2g \overline g^2 }{\omega_1-\omega_2}
\left\{-C^\mu(q_2,q_1)\frac{\Gamma^2(\epsilon)}
{\Gamma(2\epsilon)}(\vec k^2)^{\epsilon}
+\frac{(\vec q_1\vec q_2)C^\mu(q_2,q_1)+2\vec q_1^{\:2}
\vec q_2^{\:2}{\cal P}^{\mu}}{\vec q_1^{\:2}\vec q_2^{\:2}
-(\vec q_1\vec q_2)^2}
\right.
\]
\begin{equation}
\left.\times \left[\vec q_1^{\:2}\vec q_2^{\:2}\vec k^2
{\cal I}_3+\frac{\Gamma^2(\epsilon)}{\Gamma(2\epsilon)}
\left((\vec q_1^{\:2})^{\epsilon}(\vec q_2\vec k) 
-(\vec q_2^{\:2})^{\epsilon}(\vec q_1\vec k) 
-(\vec k^2)^{\epsilon}(\vec q_1\vec q_2)\right)\right]\right\}\;.
\label{34}
\end{equation} 
Evidently, the same result is obtained from the imaginary part 
of~(\ref{32}) if the representation~(\ref{26}) for ${\cal A}^\mu$ 
is used.  

The integral ${\cal I}_3$ (see~(\ref{15})) cannot be expressed through 
elementary functions at arbitrary $D$. For $\epsilon \rightarrow 0$ we 
have (see~\cite{FL93} or Appendix~C)
\begin{equation}
{\cal I}_3\simeq\frac{1}{\epsilon}\left[
\frac{(\vec q_1^{\:2}\vec q_2^{\:2})^{\epsilon-1}}
{(\vec k^2)^{\epsilon}}+
\frac{(\vec q_1^{\:2}\vec k^2)^{\epsilon-1}}
{(\vec q_2^{\:2})^{\epsilon}}+
\frac{(\vec q_2^{\:2}\vec k^2)^{\epsilon-1}}
{(\vec q_1^{\:2})^{\epsilon}}\right].
\label{35}
\end{equation} 
Using this expression we obtain from~(\ref{34})
\begin{equation}
R^\mu - L^\mu \simeq  -\frac{4g \overline g^2 }{\omega_1-\omega_2}
C^\mu(q_2,q_1)\left(\frac{1}{\epsilon}+(\vec k^2)^{\epsilon}\right)
\label{36}
\end{equation}
for $\epsilon \rightarrow 0$, in agreement with \cite{FL93} 
(see also~\cite{rio98}). 

It's easy to obtain from~(\ref{34}) also the limit $\vec k 
\rightarrow 0$ at arbitrary $D$. In this limit the main contribution
to the integral ${\cal I}_3$ (see~(\ref{15})) comes from the region 
$\vec k_1\simeq \vec q_1\simeq \vec q_2$, so that in the integrand 
the replacement 
\begin{equation}
\frac{1}{\vec k_1^{\:2}} \rightarrow  \frac{1}{\vec Q^{\:2}}\;,
\;\;\;\;\;\;\;\;\;\; \vec Q = \frac{\vec q_1+\vec q_2}{2} 
\label{36a}
\end{equation}
can be made. After this we have gently 
\begin{equation}
{\cal I}_3\simeq \frac{\Gamma^2(\epsilon)}
{\Gamma(2\epsilon)}\frac{(\vec k^2)^{\epsilon-1}}{\vec Q^{\:2}}, 
\label{36b}   
\end{equation}
and using this result we see that only the first term in curly 
brackets in~(\ref{34}) does contribute, so that in the limit $\vec k 
\rightarrow 0$ at arbitrary $\epsilon$ we obtain 
\begin{equation}
R^\mu - L^\mu = - \frac{2g \overline g^2 }{\omega_1-\omega_2}
C^\mu(q_2,q_1)\frac{\Gamma^2(\epsilon)}
{\Gamma(2\epsilon)}(\vec k^2)^{\epsilon}, 
\label{36c}   
\end{equation}
in agreement with \cite{FFK96b, DDS}. 

The real parts of~(\ref{32}), with account of~(\ref{16a}), give
\[
R^\mu + L^\mu = 2g\left[C^\mu(q_2,q_1)\left(1+
\overline g ^2 \bar r\right)\right.
\]
\begin{equation}
\left.+\overline g ^2 \left\{ - 2 t_1 t_2 ({\cal I}_5^\mu+{\cal L}_3^\mu)
 + 2 t_1 \bar r_A \frac{p_A^\mu}{s_1}
- 2 t_2 \bar r_B \frac{p_B^\mu}{s_2}\right\}\right]\;, 
\label{37}
\end{equation}
where ${\cal I}_5^\mu$ and  ${\cal L}_3^\mu$
differ from the integrals ${\cal I}_5$ and ${\cal L}_3$ defined 
in~(\ref{15}) by the factors $(k_1-q_1)_\perp^\mu$ and ${k_1}_{\perp}^\mu$, 
respectively, in the corresponding integrands, 
\[
\bar r =\left\{\frac{\vec q_1^{\:2}\vec q_2^{\:2}}{2}
({\cal I}_5-{\cal L}_3)-\vec q_2^{\:2}{\cal I}_{4B}
+\frac{\Gamma^2(\epsilon)}{2\Gamma(2\epsilon)}
(\vec q_1^{\:2})^{\epsilon}\left[\ln \left(\frac{\vec q_1^{\:2}}
{\vec k^2}\right)\!+\!\psi(1)\!+\!2\psi(\epsilon)\!-\!\psi(1-\epsilon)
\!-\!2\psi(2\epsilon)
\right.\right.
\]
\begin{equation}
\left.\left.+\frac{1}{2(1+2\epsilon)(3+2\epsilon)}\left(\frac
{\vec q_1^{\:2}+\vec q_2^{\:2}}{\vec q_1^{\:2}-\vec q_2^{\:2}}
(11+7\epsilon)+2\epsilon\frac
{\vec k^2\vec q_1^{\:2}}{(\vec q_1^{\:2}-\vec q_2^{\:2})^2}
-4\frac{\vec k^2\vec q_1^{\:2}\vec q_2^{\:2}}
{(\vec q_1^{\:2}-\vec q_2^{\:2})^3}\right)\right]\right\} 
+\left\{A\longleftrightarrow B\right\}\:, 
\label{38}
\end{equation} 
\[
\bar r_A =-\vec q_1^{\:2}\vec q_2^{\:2}
({\cal I}_5-{\cal L}_3)+\vec q_2^{\:2}{\cal I}_{4B}
+\frac{\Gamma^2(\epsilon)}{\Gamma(2\epsilon)}
(\vec q_2^{\:2})^{\epsilon}\left(\ln \left(\frac{\vec k^2}
{\vec q_2^{\:2}}\right)-\psi(1)-\psi(\epsilon)+\psi(1-\epsilon)
+\psi(2\epsilon)\right)
\]
\[
-\frac{\Gamma^2(\epsilon)}{\vec q_1^{\:2}\Gamma(2\epsilon)}
\frac{1}{2(1+2\epsilon)(3+2\epsilon)}\left\{ \left[
(\vec q_1^{\:2})^{\epsilon}\left(\frac {\vec q_1^{\:2}\vec q_2^{\:2}}
{\vec q_1^{\:2}-\vec q_2^{\:2}}
(11+7\epsilon)+\epsilon\vec k^2\vec q_1^{\:2}\frac{\vec q_1^{\:2}
-\vec k^2}{(\vec q_1^{\:2}-\vec q_2^{\:2})^2}
\right.\right.\right.
\]
\begin{equation}
\left.\left.\left.
-\vec k^2\vec q_1^{\:2}\vec q_2^{\:2}
\frac{\vec q_1^{\:2}+\vec q_2^{\:2}-2\vec k^2}
{(\vec q_1^{\:2}-\vec q_2^{\:2})^3}\right)\right]
+\left[A\longleftrightarrow B\right]\right\}\;, 
\label{39}
\end{equation} 
and 
\begin{equation}
\bar r_B =\bar r_A(A\longleftrightarrow B). 
\label{40}
\end{equation}
As well as $R^\mu - L^\mu$, we can present $R^\mu + L^\mu$ 
in the explicitly gauge invariant form. It can be done using 
the real part of~(\ref{25}) to express ${\cal I}_{5}^{\mu}$
and ${\cal L}_{3}^{\mu}$ in terms of ${\cal I}_{5}$, 
${\cal L}_{3}$ and ${\cal I}_{4A,B}$. We obtain: 
\[
R^\mu + L^\mu = 2g\left\{C^\mu(q_2,q_1)+
\overline g ^2\left(\frac{(\vec q_1\vec q_2)C^\mu(q_2,q_1)+
2\vec q_1^{\:2}\vec q_2^{\:2}{\cal P}^\mu}{2(\vec q_1^{\:2}\vec q_2^{\:2}
-(\vec q_1\vec q_2)^2)}\left[
\frac{\vec q_1^{\:2}\vec q_2^{\:2}\vec k^2}{2}
({\cal I}_5-{\cal L}_3)
\right.\right.\right.
\]
\[
\left.
-\vec q_2^{\:2}(\vec q_1\vec k)\,{\cal I}_{4B}
+\frac{\Gamma^2(\epsilon)}{\Gamma(2\epsilon)}
\left(\frac{(\vec k^2)^{\epsilon}}{2}(\vec q_1\vec q_2)\left(
\psi(\epsilon)-\psi(1-\epsilon)\right)+(\vec q_1^{\:2})^{\epsilon}
(\vec q_2\vec k)\left(\ln \left(\frac{\vec k^2}
{\vec q_1^{\:2}}\right)-\psi(1)\right.\right.\right.
\]
\[
\left.\left.\left. -\psi(\epsilon) +\psi(1-\epsilon)
+\psi(2\epsilon)\phantom{\frac{1}{1}}\!\!\!
\right)\right)\right]+C^\mu(q_2,q_1)\left[-\frac{\vec q_2^{\:2}}{2}{\cal I}_{4B}+
\frac{\Gamma^2(\epsilon)}{2\Gamma(2\epsilon)}\left(
\frac{(\vec k^2)^\epsilon}{2}(\psi(\epsilon)-\psi(1-\epsilon))
\right.\right.
\]
\[
+(\vec q_1^{\:2})^{\epsilon}\left[\psi(\epsilon)
-\psi(2\epsilon)+\frac{1}{2(1+2\epsilon)(3+2\epsilon)}\left(\frac
{\vec q_1^{\:2}+\vec q_2^{\:2}}{\vec q_1^{\:2}-\vec q_2^{\:2}}
(11+7\epsilon)+2\epsilon\frac
{\vec k^2\vec q_1^{\:2}}{(\vec q_1^{\:2}-\vec q_2^{\:2})^2} \right.\right.
\]
\[
\left.\left.\left.\left.
-4\frac{\vec k^2\vec q_1^{\:2}\vec q_2^{\:2}}
{(\vec q_1^{\:2}-\vec q_2^{\:2})^3}\right)\right]\right)\right]
+{\cal P}^\mu\frac{\Gamma^2(\epsilon)(\vec q_1^{\:2})^{\epsilon}}
{\Gamma(2\epsilon)(1+2\epsilon)(3+2\epsilon)}
\left[\frac{\vec q_1^{\:2}\vec q_2^{\:2}}{\vec q_1^{\:2}-\vec q_2^{\:2}}
(11+7\epsilon)+\epsilon\vec k^2\vec q_1^{\:2}\frac
{\vec q_1^{\:2}-\vec k^2}{(\vec q_1^{\:2}-\vec q_2^{\:2})^2}
\right.
\]
\begin{equation}
\left.\left.\left.
-\vec k^2\vec q_1^{\:2}\vec q_2^{\:2}
\frac{\vec q_1^{\:2}+\vec q_2^{\:2}-2\vec k^2}
{(\vec q_1^{\:2}-\vec q_2^{\:2})^3}\right]\right)
-\overline g ^2\left(A\longleftrightarrow B\right)\right\}\;.  
\label{41}
\end{equation}
The same result is obtained from the real part 
of~(\ref{32}) if for ${\cal A}^\mu$ the representation~(\ref{26}) 
is used.  

The integrals ${\cal I}_{5}$, ${\cal L}_{3}$ and ${\cal I}_{4A,B}$ 
defined in~(\ref{15}) cannot be written in terms of elementary functions at 
arbitrary $\epsilon$. For $\epsilon \rightarrow 0$ we have, with 
accuracy up to terms vanishing in this limit (see Appendix~C):
\[
{\cal I}_5-{\cal L}_3\simeq-\frac{1}{\vec q_1^{\:2}\vec q_2^{\:2}}
\left[\frac{1}{\epsilon^2}
\left(\frac{\vec q_1^{\:2}\vec q_2^{\:2}}
{\vec k^2}\right)^\epsilon-\frac{\pi^2}{6}\right]
\]
\begin{equation}
+\frac{1}{\vec q_1^{\:2}\vec k^2}
\left[\frac{1}{\epsilon^2}
\left(\frac{\vec q_1^{\:2}\vec k^2}
{\vec q_2^{\:2}}\right)^\epsilon-\frac{\pi^2}{2}
-2L\left(1-\frac{\vec q_1^{\:2}}{\vec q_2^{\:2}}\right)\right]+
\frac{1}{\vec q_2^{\:2}\vec k^2}
\left[\frac{1}{\epsilon^2}
\left(\frac{\vec q_2^{\:2}\vec k^2}
{\vec q_1^{\:2}}\right)^\epsilon-\frac{\pi^2}{2}
-2L\left(1-\frac{\vec q_2^{\:2}}{\vec q_1^{\:2}}\right)\right]\;,
\label{42}
\end{equation} 
\begin{equation}
{\cal I}_{4B}\simeq \frac{1}{\vec q_2^{\:2}}
\left[\frac{(\vec q_2^{\:2})^\epsilon-2(\vec q_1^{\:2})^\epsilon}{\epsilon^2}
-\frac{\pi^2}{6}+2L\left(1-\frac{\vec q_1^{\:2}}{\vec q_2^{\:2}}\right)\right]\;,
\label{43}
\end{equation}
where 
\begin{equation}
L(x)=\int_0^x\frac{dt}{t}\ln (1-t)\;.
\label{44}
\end{equation}
Using these integrals, we obtain from~(\ref{41})
\[
R^\mu + L^\mu = 2g\left\{C^\mu(q_2,q_1)\left(1+
\overline g ^2\left[-\frac{1}{\epsilon^2}-\frac{1}{\epsilon}
\ln (\vec k^2)-\frac{1}{2}\ln ^2(\vec k^2)- \frac{1}{2}\ln ^2
\left(\frac{\vec q_1^{\:2}}{\vec q_2^{\:2}}\right)+\frac{\pi^2}{2}
\right.\right.\right.
\]
\[
\left.\left.
+ \frac{\vec k^2}{3}\frac{\vec q_1^{\:2}+\vec q_2^{\:2}}
{(\vec q_1^{\:2}-\vec q_2^{\:2})^2}
+\frac{1}{6}\left(11
\frac{\vec q_1^{\:2}+\vec q_2^{\:2}}
{\vec q_1^{\:2}-\vec q_2^{\:2}} -\frac{4\vec k^2\vec q_1^{\:2}
\vec q_2^{\:2}}{(\vec q_1^{\:2}-\vec q_2^{\:2})^3}\right)\ln 
\left(\frac{\vec q_1^{\:2}}{\vec q_2^{\:2}}\right)\right]\right)
\]
\[
+{\cal P}^\mu \overline g^2\frac{2}{3}\left[
\frac{\vec k^2}{(\vec q_1^{\:2}-\vec q_2^{\:2})^2}
\left(\vec q_1^{\:2}(\vec q_1^{\:2}-\vec k^2)
+     \vec q_2^{\:2}(\vec q_2^{\:2}-\vec k^2)\right)
+\left(11\frac{\vec q_1^{\:2}\vec q_2^{\:2}}
{\vec q_1^{\:2}-\vec q_2^{\:2}} \right.\right.
\]
\begin{equation}
\left.\left.\left.
-\frac{\vec k^2\vec q_1^{\:2}
\vec q_2^{\:2}}{(\vec q_1^{\:2}-\vec q_2^{\:2})^3}
(\vec q_1^{\:2}+\vec q_1^{\:2}-2\vec k^2)\right)\ln 
\left(\frac{\vec q_1^{\:2}}{\vec q_2^{\:2}}\right)\right]\right\}\;,
\label{45}
\end{equation}
in agreement with~\cite{FL93}, taking into account the difference in 
the energy scales (see~\cite{rio98}) and the charge renormalization 
(remind that in all formulae above the bare charge $g$ is used).

The limit $\vec k \rightarrow 0$ at arbitrary $D$ can be also  
considered without difficulties. The value of the integral 
$I_5$ in this limit is known~\cite{FFK96b}. In the $s$-channel 
physical region  
\begin{equation}
I_5 \simeq \pi^{2+\epsilon}\Gamma(1-\epsilon) \frac{\Gamma^2(\epsilon)}
{\Gamma(2\epsilon)}\frac{(\vec k^2)^{\epsilon-1}}{s\vec Q^{\:2}}
\left[\psi(\epsilon)-\psi(1-\epsilon)+i\pi\right]. 
\label{45a}   
\end{equation}
Therefore from~(\ref{C2}) and~(\ref{36b}) we obtain 
\begin{equation}
{\cal I}_5-{\cal L}_3 \simeq \frac{\Gamma^2(\epsilon)}
{\Gamma(2\epsilon)}\frac{(\vec k^2)^{\epsilon-1}}{\vec Q^{\:2}}
\left[\psi(1-\epsilon)-\psi(\epsilon)\right].  
\label{46a}   
\end{equation}
The integral ${\cal I}_{4A}$ (see~(\ref{15})) is finite at $\vec k 
\rightarrow 0$ and can  be 
calculated in this limit by standard methods. We have 
\begin{equation}
{\cal I}_{4A} \simeq \frac{\Gamma^2(\epsilon)}
{\Gamma(2\epsilon)}(\vec Q^2)^{\epsilon-1}
\left[\psi(\epsilon)-\psi(2\epsilon)\right].  
\label{47a}   
\end{equation}
Using~(\ref{46a}) and~(\ref{47a}), we obtain from~(\ref{41}) 
\begin{equation}
R^\mu + L^\mu = 2g C^\mu(q_2,q_1)\left(1+\overline g^2
\frac{\Gamma^2(\epsilon)}{2\Gamma(2\epsilon)}(\vec k^2)^{\epsilon}
\left[\psi(\epsilon)-\psi(1-\epsilon)\right]\right), 
\label{48b}   
\end{equation}
that agrees  with \cite{FFK96b, DDS}, taking into account the 
difference in the energy scales (see~\cite{rio98}).  

\section{The one-gluon contribution to the BFKL kernel} 
\setcounter{equation}{0} 

The complicated analytical structure~(\ref{28}) of the RRG-vertex is
irrelevant for the calculation of the contribution of the 
one-gluon production to the BFKL kernel at the NLO,  
where only the real parts of the production amplitudes do contribute, because 
only these parts interfere with the leading order amplitudes, which are real. 
Therefore, with the required accuracy, the production amplitude 
can be presented in the same form as in the leading order: 
\begin{equation}
A_{2\rightarrow 3} = 2s \, \Gamma_{A'A}^{c_1} \, 
\left(\frac{s_1}{\vec k^2}\right)^{\omega_1}\frac{1}{t_1} \, 
\,T_{A'A}^{c_1} \, \gamma_{c_1c_2}^{d}(q_1,q_2) 
\, \left(\frac{s_2}{\vec k^2}\right)^{\omega_2}\frac{1}{t_2} \, 
\,T_{B'B}^{c_2} \;,
\label{46}
\end{equation}
where $\omega_i\equiv \omega(t_i)$, $\vec k^{\:2}$ is used as the 
energy scale and 
\begin{equation} 
\gamma_{c_1c_2}^{d}(q_1,q_2)=\frac{1}{2}\, T_{c_2c_1}^{d}\,e^*_\mu(k) \, 
(R^\mu + L^\mu) \;.  
\label{47}
\end{equation}
The  contribution of the one-gluon production to the BFKL kernel 
for non-forward scattering with momentum transfer $q$ and irreducible 
representation $\cal R$ of the colour group in the $t$-channel 
is~\cite{rio98,FF98} 
\begin{equation}
{\cal K}_{RRG}^{G\,({\cal R})}(\vec q_1,\vec q_2; \vec{q})=
\frac{1}{(2\pi)^{D-1}}\frac{\langle c_1c_1^{\prime }|\hat {{\cal
P}}_{ {\cal R}}|c_2c_2^{\prime }\rangle }{2n_{{\cal
R}}}\sum_{d,\lambda}\gamma _{c_1c_2}^{d}\left( q_1,q_2\right)
\left( \gamma _{c_1^{\prime}c_2^{\prime }}^{d}\left( q_1^{\prime
},q_2^{\prime }\right) \right) ^{*}\;,   
\label{48}
\end{equation}
where $n_{{\cal R}}$ is the number of independent states in the
representation ${\cal R}$, the sum is performed over colours and 
polarizations of the produced gluons and 
\begin{equation}
q_1^{\prime}=q_1- q\;,\;\;\;\;\;\;\;\;\;\;\; q_2^{\prime}=q_2-q\,.
\label{48a}
\end{equation}  
The most interesting representations ${\cal R}$ are the colour
singlet (Pomeron channel) and the antisymmetric colour octet
(gluon channel). We have for the singlet case
\begin{equation}
\langle c_1c_1^{\prime }|\hat {{\cal P}}_1|c_2c_2^{\prime
}\rangle =\frac{ \delta _{c_1c_1^{\prime }}\delta
_{c_2c_2^{\prime }}}{N^2-1}\;, \;\;\;\;\;\;\;\;\;\; n_1=1\;, \label{49}
\end{equation}
and for the octet case
\begin{equation}
\langle c_1c_1^{\prime }|\hat {{\cal P}}_8|c_2c_2^{\prime
}\rangle =\frac{ f_{c_1c_1^{\prime }c}f_{c_2c_2^{\prime }c}}N\;,
\;\;\;\;\;\;\;\;\;\; n_8=N^2-1\;, \label{50}
\end{equation}
where $f_{abc}$ are the structure constants of the colour group.
The vertex~(\ref{47}) is explicitly invariant under the gauge
transformation
\begin{equation}
e^\mu (k)\rightarrow e^\mu (k)+k^\mu \chi\;, \label{51}
\end{equation}
so that we can use the relation
\begin{equation}
\sum_{\lambda}e^{*\:(\lambda)}_{\mu}(k)e_{\nu}^{(\lambda)}(k)=-g_{\mu \nu}\;.
\label{52}
\end{equation}
Substituting~(\ref{47}) in~(\ref{48}), using~(\ref{52}) for
the sum over polarizations and
\begin{equation}
\delta_{c_1c_1'}\delta_{c_2c_2'}T_{c_1c_2}^{d}\left(T_{c_1'c_2'}^{d}
\right)^*\hspace{-1mm}
=N(N^2-1)\:,\;\;\;\;\;\;\;\;
f_{c_1c_1'c}f_{c_2c_2'c}T_{c_1c_2}^{d}\left(T_{c_1'c_2'}^{d}\right)^*
\hspace{-1mm}
=\frac{N^2(N^2-1)}{2} \label{53}
\end{equation}
for the sum over colour indices, we obtain with the NLO accuracy: 
\[
{\cal K}_{RRG}^{G\,({\cal R})}(\vec q_1,\vec q_2; \vec{q})
=  -\frac{c_{\cal 
R}}{8(2\pi)^{D-1}}2g\left[C_\mu(q_2^{\prime},q_1^{\prime})
(R+L)^\mu +C_\mu(q_2,q_1)(R'+L')^\mu 
\right.
\]
\begin{equation}
\left.-2gC_\mu(q_2,q_1)
C^\mu(q_2^{\prime},q_1^{\prime})\right]\;,
\label{54}
\end{equation}
where for the singlet (${\cal R}=1$) and octet (${\cal R}=8$) cases 
\begin{equation}
c_1=N\;,\;\;\;\;\;\;\;\;\;\;\; c_8=\frac{N}{2}\;,
\label{55}
\end{equation}        
$C_\mu(q_2,q_1)$ is defined in~(\ref{7}), the sum $(R+L)^\mu$ 
is presented in two different forms in~(\ref{37})-(\ref{40}) 
and~(\ref{41}), $(R'+L')^\mu$ is obtained from $(R+L)^\mu$ by the 
substitution 
\begin{equation}
\vec q_1\rightarrow \vec q_1^{\:\prime}=\vec q_1-\vec q\;,\;\;\;\;\;\;\;\;\; 
\vec q_2\rightarrow \vec q_2^{\:\prime}=\vec q_2-\vec q\,.
\label{56}
\end{equation}  
The convolution over the vector indices in~(\ref{54}) is easily performed 
with the help of the relations 
\[
C_\mu(q_2,q_1)C^\mu(q_2^{\prime},q_1^{\prime})=
2 \left(\vec{q}^{\:2}- \frac{\vec{q}_1^{\:2} \vec{q}_2^{\:\prime\:2}+
\vec{q}_1^{\:\prime\:2} \vec{q}_2^{\:2}}{\vec k^2}\right)\:,
\]
\begin{equation}
C_\mu(q_2,q_1){\cal P}^\mu = \frac{\vec{q}_1^{\:2}
+\vec{q}_2^{\:2}}{\vec k^2}-1\;. 
\label{57}
\end{equation}
In general case the convolution does not lead to noticeable 
simplifications, so that it has no sense to rewrite~(\ref{54}) 
in unfolded form. But at $\vec q=0$ there is a huge simplification 
due to the equality 
\begin{equation}
C_\mu(q_2,q_1)\left[(\vec{q}_1 \vec{q}_2)C^\mu(q_2,q_1)
+2\vec{q}_1^{\:2} \vec{q}_2^{\:2}{\cal P}^\mu\right]  = 0\;.
\label{58}
\end{equation} 
For this case we obtain  
\[
{\cal K}_{RRG}^{G\,({\cal R})}(\vec q_1,\vec q_2; \vec 0)=
\frac{g^2c_{\cal R}}{(2\pi)^{D-1}}\left\{\frac{\vec q_1^{\:2}\vec 
q_2^{\:2}}
{\vec k^2}\left[1+2{\overline g^2}\left(-\vec q_2^{\:2}{\cal 
I}_{4B}+
\frac{\Gamma^2(\epsilon)}{\Gamma(2\epsilon)}\left(
\frac{(\vec k^2)^\epsilon}{2}(\psi(\epsilon)-\psi(1-\epsilon))
\right.\right.\right.\right.
\]
\[
\left.\left.\left.
+(\vec q_1^{\:2})^\epsilon(\psi(\epsilon)
-\psi(2\epsilon))\phantom{\frac{1}{1}}\!\!\!\!
\right)\right)\right]-\frac{\Gamma^2(\epsilon)}
{\Gamma(2\epsilon)}\frac{{\overline g^2}(\vec 
q_1^{\:2})^{\epsilon+1}}
{(1+2\epsilon)(3+2\epsilon)}\left[-12\frac{\vec q_2^{\:2}}
{\vec q_1^{\:2}-\vec q_2^{\:2}} \right.
\]
\[
\left.\left.
+\frac{\vec k^2\vec q_2^{\:2}}
{(\vec q_1^{\:2}-\vec q_2^{\:2})^3}(3(\vec q_1^{\:2}+\vec q_2^{\:2})
-2\vec k^2)
+\epsilon\left(\frac{\vec q_1^{\:2}-7\vec q_2^{\:2}}
{(\vec q_1^{\:2}-\vec q_2^{\:2})}-\frac{\vec k^2}
{(\vec q_1^{\:2}-\vec q_2^{\:2})^2}(2\vec q_1^{\:2}+\vec q_2^{\:2}-
\vec k^2)\right) \right]\right\}
\]
\begin{equation}
+\frac{g^2c_{\cal R}}{(2\pi)^{D-1}}\biggl\{A\longleftrightarrow B\biggr\}\;.
\label{59}
\end{equation}           

\vspace{1.0cm} 
\underline{Acknowledgment}: V.S.F. thanks the Dipartimento di Fisica della 
Universit\`a della Calabria and the Istituto Nazionale di Fisica Nucleare - 
Gruppo collegato di Cosenza for the warm hospitality while part of this 
work was done. 
 
\appendix

\section{Appendix}

In this Appendix we give the expressions of the integrals relevant for this paper, 
among those having the general form of Eqs.~(\ref{14}). The integrals 
with two or three denominators were already calculated in~\cite{FL93} 
at arbitrary $D$. Those with four and five denominators were calculated in 
Ref.~\cite{FL93} in the $\epsilon$-expansion. Unfortunately, it is not possible 
in general to express them in terms of elementary functions at arbitrary $D$.
It is possible, however, to perform the integration over the longitudinal variables
in the Sudakov decomposition for the integration variable $p$ and to express 
the result in terms of $(D-2)$-dimensional integrals over $\vec p$. As an 
illustration of the method, we will show in some details the steps involved 
in the calculation of $I_4^{(1)}$ for which it turns out that, in the 
multi-Regge kinematics~(\ref{2}), the complete integration can be performed. 
For the remaining integrals with four of five denominators, we will merely 
list the results.

Let us start from the integrals with two or three denominators:
\begin{equation}
I_2(q) = \frac{1}{i} \int \, d^D p \, \frac{1}{(p^2+i\varepsilon)
[(p+q)^2+i\varepsilon]} 
= -\frac{\pi^{2+\epsilon}\Gamma(1-\epsilon)}{2(1+2\epsilon)}\, 
\frac{\Gamma^2(\epsilon)}{\Gamma(2\epsilon)}(\vec q^{\:2})^\epsilon\;,
\label{I2}
\end{equation}
\begin{equation}
I_2^{\mu\nu}(q) = \frac{1}{i} \int \, d^D p \, \frac{p^\mu p^\nu}
{(p^2+i\varepsilon)[(p+q)^2+i\varepsilon]} 
= \frac{1}{4(3+2\epsilon)}\, \biggl[2(2+\epsilon)\,q^\mu q^\nu - q^2\, g^{\mu\nu}
\biggr] I_2(q)\;, 
\label{I2^m}
\end{equation}
\begin{equation}
I_3 = \frac{1}{i} \int \, d^D p \, \frac{1}
{(p^2+i\varepsilon)[(p+q_1)^2+i\varepsilon] [(p+k)^2+i\varepsilon]} 
= \frac{1+2\epsilon}{\epsilon} \, \frac{I_2(q_2)-I_2(q_1)}{t_1-t_2}\;,
\label{I3}
\end{equation}
\[
I_3^\mu = \frac{1}{i} \int \, d^D p \, \frac{p^\mu}
{(p^2+i\varepsilon)[(p+q_1)^2+i\varepsilon] [(p+k)^2+i\varepsilon]} 
= q_1^\mu \, \frac{I_2(q_1)-I_2(q_2)}{t_1-t_2} 
\]
\begin{equation}
+ \frac{k^\mu}{(t_1-t_2)^2} \, \left[\frac{t_1}{\epsilon} I_2(q_1)
+\left(t_2-\frac{1+\epsilon}{\epsilon}\,t_1\right) I_2(q_2)\right]\;,
\label{I3^m}
\end{equation}
\[
I_3^{\mu\nu} = \frac{1}{i} \int \, d^D p \, \frac{p^\mu p^\nu}
{(p^2+i\varepsilon)[(p+q_1)^2+i\varepsilon] [(p+k)^2+i\varepsilon]} 
= \frac{g^{\mu\nu}}{4(1+\epsilon)}\frac{t_1 I_2(q_1)-t_2 I_2(q_2)}{t_1-t_2}
\]
\[ 
-\frac{q_1^\mu q_1^\nu}{2}\frac{I_2(q_1)-I_2(q_2)}{t_1-t_2}
+\frac{k^\mu q_1^\nu+k^\nu q_1^\mu}{2(1+\epsilon)}\,
\frac{-t_1 I_2(q_1) + [(1+\epsilon)\,t_1-\epsilon\,t_2)] I_2(q_2)}
{(t_1-t_2)^2}
\]
\begin{equation}
+\frac{k^\mu k^\nu}{(t_1-t_2)^3} \,\left[-\frac{t_1^2 I_2(q_1)}
{\epsilon(1+\epsilon)} + \left(\frac{t_1^2}{\epsilon}
-\frac{t_1 t_2}{1+\epsilon}+\frac{(t_1-t_2)^2}{2}\right) I_2(q_2)\right]\;,
\label{I3^mn}
\end{equation}
\[
I_3^{\mu\nu\rho} = \frac{1}{i} \int \, d^D p \, \frac{p^\mu p^\nu p^\rho}
{(p^2+i\varepsilon)[(p+q_1)^2+i\varepsilon] [(p+k)^2+i\varepsilon]} 
\]
\[
= \biggl(g^{\mu\nu}q_1^\rho+g^{\mu\rho}q_1^\nu+g^{\nu\rho}q_1^\mu\biggr)
\frac{t_2 I_2(q_2)-t_1 I_2(q_1)}{4(3+2\epsilon)(t_1-t_2)}
\]
\[
-\biggl(g^{\mu\nu}k^\rho+g^{\mu\rho}k^\nu+g^{\nu\rho}k^\mu\biggr)
\frac{t_1^2 I_2(q_1)-[(2+\epsilon)t_2(t_1-t_2)+t_2^2] I_2(q_2)}
{4(1+\epsilon)(3+2\epsilon)(t_1-t_2)^2}
\]
\[
+q_1^\mu q_1^\nu q_1^\rho \frac{(2+\epsilon)\,[I_2(q_1)-I_2(q_2)]}
{2(3+2\epsilon)(t_1-t_2)}
\]
\[
+\biggl(q_1^\mu q_1^\nu k^\rho + q_1^\mu k^\nu q_1^\rho 
+ k^\mu q_1^\nu q_1^\rho\biggr)
\frac{t_1 I_2(q_1)-[t_2+(1+\epsilon)(t_1-t_2)] I_2(q_2)}
{2(3+2\epsilon)(t_1-t_2)^2}
\]
\[
+\biggl(q_1^\mu k^\nu k^\rho + k^\mu q_1^\nu k^\rho 
+ k^\mu k^\nu q_1^\rho\biggr)
\frac{2 t_1^2 I_2(q_1)\!-\![2 t_2^2\!+\!(2+\epsilon)(t_1-t_2)(2 t_2
+(1+\epsilon)(t_1-t_2))] I_2(q_2)}{2(1+\epsilon)(3+2\epsilon)(t_1-t_2)^3}
\]
\[
+k^\mu k^\nu k^\rho \left\{\frac{3[t_1^3 I_2(q_1)-t_2^3 I_2(q_2)]}
{\epsilon (1+\epsilon)(3+2\epsilon)(t_1-t_2)^4} \right.
\]
\[
-\left[\frac{3(3+\epsilon)t_2^2}{\epsilon (1+\epsilon) (3+2\epsilon)(t_1-t_2)^3}
+\frac{3 t_2}{(t_1-t_2)^2}\left(\frac{1}{2(3+2\epsilon)}
+\frac{1}{\epsilon(1+\epsilon)}\right)\right.
\]
\begin{equation}
\left. \left.
+\frac{1}{t_1-t_2}\left(-\frac{1}{4(3+2\epsilon)}+\frac{1}{\epsilon}
+\frac{1}{4}\right) \right] I_2(q_2)\right\}\;, 
\label{I3^mnr}
\end{equation}
\begin{equation}
I_{3A} = \frac{1}{i} \int \, d^D p \, \frac{1}
{(p^2+i\varepsilon)[(p+q_1)^2+i\varepsilon] [(p+p_A)^2+i\varepsilon]} 
= -\frac{1+2\epsilon}{\epsilon} \frac{I_2(q_1)}{t_1} \;,
\label{I3A}
\end{equation}
\begin{equation}
I_{3A}^\mu = \frac{1}{i} \int \, d^D p \, \frac{p^\mu}
{(p^2+i\varepsilon)[(p+q_1)^2+i\varepsilon] [(p+p_A)^2+i\varepsilon]} 
=\left(q_1^\mu+\frac{p_A^\mu}{\epsilon}\right)\frac{I_2(q_1)}{t_1}\;.
\label{I3A^m}
\end{equation}

Let us consider now integrals with four or five denominators. As anticipated
above, we will show in some detail how to calculate $I_4^{(1)}$ and simply 
give the final results for the remaining integrals.

The integral under consideration is 
\begin{equation}
I_4^{(1)} = \frac{1}{i} \int \, d^D p \, \frac{1}
{(p^2+i\varepsilon)[(p+q_1)^2+i\varepsilon][(p+p_A)^2+i\varepsilon]
[(p-p_B)^2+i\varepsilon]} \;.
\end{equation}
The Sudakov decompositions for the integration momentum $p$ and 
for $q_1$ are
\begin{equation}
p = \beta p_A + \alpha p_B + p_\perp\;, \;\;\;\;\;\;\;\;\;\;
q_1 \equiv p_A - p_{A^\prime} = \frac{s_2}{s} p_A - \frac{\vec q_1^{\:2}}{s} p_B 
+ {q_1}_\perp \;.
\end{equation}
After changing the integration variables to $\alpha$, $\beta$ and $p_\perp$ and
using
\begin{equation}
d^D p = \frac{s}{2} \, d\alpha \, d\beta \, d^{D-2}p\;,
\end{equation}
we perform the integration over the variable $\alpha$ by the method
of residues. This leads in the multi-Regge kinematics to the following 
result
\[
I_4^{(1)} = \pi \int_0^1 d\beta \beta^2 \int \frac{d^{D-2}p}{\vec p^{\:2} 
(\vec p +\beta \vec q_1)^2 [\vec p^{\:2}+\beta(1-\beta)(-s-i\varepsilon)]}
\]
\begin{equation}
- \pi \frac{s_2}{s} \int_0^1 d\beta \beta^2 \int \frac{d^{D-2}p}
{\vec p^{\:2} [(\vec p + \beta \vec q_1)^2+\beta(1-\beta)\vec q_1^{\:2}] 
[(\vec p + \beta \vec q_1)^2+\beta(1-\beta)(-s_2-i\varepsilon)]}\;.
\label{I4_1}
\end{equation}
The first integral in the R.H.S. of the above equation can be manipulated as
follows:
\[
\biggl[I_4^{(1)}\biggr]_1 \equiv 
\pi\int_0^1 d\beta \beta^2 \int \frac{d^{D-2}p}{\vec p^{\:2} 
(\vec p +\beta \vec q_1)^2 [\vec p^{\:2}+\beta(1-\beta)(-s-i\varepsilon)]}
\]
\[
= \pi\int \frac{d^{D-2}p}{\vec p^{\:2} (\vec p + \vec q_1)^2}
\int_0^1 d\beta  \frac{(1-\beta)^{D-5}}{(1-\beta) \vec p^{\:2} 
+\beta(-s-i\varepsilon)}
\]
\begin{equation}
\simeq \pi\int \frac{d^{D-2} p}{\vec p^{\:2} (\vec p + \vec q_1)^2}
\int_0^1 d\beta  \frac{(1-\beta)^{D-5}}{\vec p^{\:2} +\beta(-s-i\varepsilon)}\;,
\label{I4_1_1_in}
\end{equation}
where the first equality follows from the change of variables 
$\vec p \rightarrow \beta \vec p$ and $\beta \rightarrow 1-\beta $,
while
the last approximated equality holds since in the $s\rightarrow
 \infty$ limit $s+\vec p^{\:2} = s$.
Then, the integral over $\beta$ can be performed:
\[
\int_0^1 d\beta \frac{(1-\beta)^{D-5}}{\delta + \beta}
= \int_0^1 d\beta \frac{(1-\beta)^{D-5}-1}{\delta + \beta}
+ \int_0^1 \frac{d\beta}{\delta + \beta}
\]
\begin{equation}
\simeq \int_0^1 d\beta \frac{(1-\beta)^{D-5}-1}{\beta}
+ \int_0^1 \frac{d\beta}{\delta + \beta}
\simeq \psi(1)-\psi(D-4)+\ln\left(\frac{1}{\delta}\right) \;,
\end{equation}
where $\delta \equiv \vec p^{\:2}/(-s-i\varepsilon)$ is a quantity tending to zero
and $\psi(x)$ is the logarithmic derivative of $\Gamma(x)$.
After performing the integration over $p_\perp$, we obtain for 
$\biggl[I_4^{(1)}\biggr]_1$ the following result:
\begin{equation}
\biggl[I_4^{(1)}\biggr]_1 = \frac{\pi^{2+\epsilon}\Gamma(1-\epsilon)}{s \, t_1}
\frac{\Gamma^2(\epsilon)}{\Gamma(2\epsilon)}\,(\vec q_1^{\:2})^\epsilon\,
\left[\ln\left(\frac{-s-i\varepsilon}{-t_1}
\right) +\psi(1-\epsilon) -\psi(\epsilon)\right]\;.
\label{I4_1_1}
\end{equation}
In the last expression, the terms $-i\varepsilon$ which appears in the argument 
of the logarithm fixes the prescription for the analytic
continuation of this function to the region of positive $s$. In the following,
this term will always be omitted in the final results, but it should be 
understood to accompany $-s$, $-s_1$ or $-s_2$ every time these
quantities appear in the argument of the logarithm.
The second integral in the R.H.S. of Eq.~(\ref{I4_1}) can be decomposed as 
follows:
\[
\biggl[I_4^{(1)}\biggr]_2 = 
- \pi \frac{s_2}{s} \int_0^1 d\beta \beta^2 \int \frac{d^{D-2}p}
{(\vec p+\beta \vec q_1)^2 [\vec p^{\:2} + \beta(1-\beta)\vec q_1^{\:2}] 
[\vec p^{\:2} + \beta(1-\beta)(-s_2-i\varepsilon)]}
\]
\[
= - \pi \frac{\vec q_1^{\:2}}{s} 
\int_0^1 d\beta \beta^2 \int \frac{d^{D-2}p}
{\vec p^{\:2}(\vec p+\beta \vec q_1)^2[\vec p^{\:2}+\beta(1-\beta)\vec q_1^{\:2}]} 
\]
\begin{equation}
- \pi \frac{s_2}{s} 
\int_0^1 d\beta \beta^2 \int \frac{d^{D-2}p}
{\vec p^{\:2}(\vec p+\beta \vec q_1)^2[\vec p^{\:2}+\beta(1-\beta)
(-s_2-i\varepsilon)]}\;,
\end{equation} 
where it has been used that in the multi-Regge kinematics $s_2\gg\vec q_1^{\:2}$.
The second integral in the R.H.S. of the above equation has the same structure
as $\biggl[I_4^{(1)}\biggr]_1$ (see Eq.~(\ref{I4_1_1_in})) and does not need
to be calculated. The first one can be easily evaluated by the usual Feynman 
parametrization technique. The final result for $\biggl[I_4^{(1)}\biggr]_2$ is 
then
\begin{equation}
\biggl[I_4^{(1)}\biggr]_2 = \frac{\pi^{2+\epsilon}\Gamma(1-\epsilon)}{s \, t_1}
\frac{\Gamma^2(\epsilon)}{\Gamma(2\epsilon)}\,(\vec q_1^{\:2})^\epsilon\,
\left[\ln\left(\frac{-t_1}{-s_2}\right) +\psi(1) -\psi(1-\epsilon)\right]\;.
\label{I4_1_2}
\end{equation}
Combining Eqs.~(\ref{I4_1_1}) and~(\ref{I4_1_2}), we obtain the final result
for $I_4^{(1)}$:
\begin{equation}
I_4^{(1)} = \frac{\pi^{2+\epsilon}\Gamma(1-\epsilon)}{s \, t_1}
\frac{\Gamma^2(\epsilon)}{\Gamma(2\epsilon)}\,(\vec q_1^{\:2})^\epsilon\,
\left[\ln\left(\frac{-s}{-s_2}\right) +\psi(1) -\psi(\epsilon)\right]\;.
\label{I4_1_fin}
\end{equation} 

The same procedure described for the calculation of $I_4^{(1)}$ can be applied
for the integrals $I_{4A}$, $I_5$ and $I_5^\mu$, except that for them
the integrations in the transverse space cannot be performed in complete way. 
We list here the final results:
\begin{equation}
I_{4A} = - \frac{\pi^{2+\epsilon}\Gamma(1-\epsilon)}{s_1}
\left[\frac{\Gamma^2(\epsilon)}{\Gamma(2\epsilon)}\,
(\vec q_1^{\:2})^{\epsilon-1}\,
\left(\ln\left(\frac{-s_1}{-t_1}\right) + \psi(1-\epsilon) -2\psi(\epsilon)
+\psi(2\epsilon)\right)+{\cal I}_{4A}\right]\;,
\label{I4A}
\end{equation}
\begin{equation}
I_5 = \frac{\pi^{2+\epsilon}\Gamma(1-\epsilon)}{s}\left[
\ln\left(\frac{(-s)\vec k^2}{(-s_1)(-s_2)}\right){\cal I}_3 
+{\cal L}_3-{\cal I}_5\right]\;,
\label{I5}
\end{equation}
\begin{equation}
I_5^\mu = q_1^\mu I_5 - p_A^\mu \frac{I_{4A}}{s} + p_B^\mu \frac{I_{4B}}{s}
+\frac{\pi^{2+\epsilon}\Gamma(1-\epsilon)}{s}
\left[-\ln\left(\frac{(-s)\vec k^2}{(-s_1)(-s_2)}\right)
{\cal I}_3^\mu 
-{\cal L}_3^\mu-{\cal I}_5^\mu\right]\;,
\label{I5^m}
\end{equation}
where 
\[
{\cal I}_{4A} \!\!=\!\! \int_0^1 \!\!\!\frac{dx}{1-x}\! \int\!\!\! 
\frac{d^{2+2\epsilon}k_1}
{\pi^{1+\epsilon} \Gamma(1-\epsilon)} \! \left[\frac{x}{[(1-x) \vec k_1^2+
x (\vec k_1+\vec q_2)^2](\vec k_1-x\vec k)^2} 
-\frac{1}{(\vec k_1+\vec q_2)^2 (\vec k_1-\vec k)^2}\right]\!,
\]
\[
{\cal I}_5 = \int_0^1 \frac{dx}{1-x} \int \frac{d^{2+2\epsilon}k_1}
{\pi^{1+\epsilon} \Gamma(1-\epsilon)} \frac{1}{\vec k_1^2 [(1-x) \vec k_1^2+
x (\vec k_1-\vec q_1)^2]} \left(\frac{x^2}{(\vec k_1-x\vec k)^2}-\frac{1}
{(\vec k_1-\vec k)^2}\right) \;,
\]
\[
{\cal I}_5^\mu = \int_0^1 \frac{dx}{1-x} \int \frac{d^{2+2\epsilon}k_1}
{\pi^{1+\epsilon} \Gamma(1-\epsilon)} \frac{(k_1-q_1)_\perp^\mu}
{\vec k_1^2 [(1-x) \vec k_1^2+
x (\vec k_1-\vec q_1)^2]} \left(\frac{x^2}{(\vec k_1-x\vec k)^2}-\frac{1}
{(\vec k_1-\vec k)^2}\right) \;,
\]
\[
{\cal L}_3 = \int \frac{d^{2+2\epsilon}k_1}
{\pi^{1+\epsilon} \Gamma(1-\epsilon)} \frac{1}{\vec k_1^2 (\vec k_1-\vec q_1)^2
(\vec k_1-\vec q_2)^2} \ln\left(\frac{(\vec k_1-\vec q_1)^2 (\vec k_1-\vec q_2)^2}
{\vec k^2 \, \vec k_1^2}\right)\;,
\]
\[
{\cal L}_3^\mu = \int \frac{d^{2+2\epsilon}k_1}
{\pi^{1+\epsilon} \Gamma(1-\epsilon)} \frac{{k_1}_\perp^\mu}
{\vec k_1^2 (\vec k_1-\vec q_1)^2
(\vec k_1-\vec q_2)^2} \ln\left(\frac{(\vec k_1-\vec q_1)^2 (\vec k_1-\vec q_2)^2}
{\vec k^2 \, \vec k_1^2}\right)\;,
\]
\[
{\cal I}_3 = \int \frac{d^{2+2\epsilon}k_1}
{\pi^{1+\epsilon} \Gamma(1-\epsilon)} \frac{1}{\vec k_1^2 (\vec k_1-\vec q_1)^2
(\vec k_1-\vec q_2)^2} \;,
\]
\begin{equation}
{\cal I}_3^\mu = \int \frac{d^{2+2\epsilon}k_1} 
{\pi^{1+\epsilon} \Gamma(1-\epsilon)} \frac{{k_1}_\perp^\mu}
{\vec k_1^2 (\vec k_1-\vec q_1)^2 (\vec k_1-\vec q_2)^2} \;,
\end{equation}
$I_{4B}$ in the expression for $I_5^\mu$ (Eq.~(\ref{I5^m})) is obtained from 
$I_{4A}$ by the replacements $p_A \longleftrightarrow p_B$ and 
$q_1 \longleftrightarrow -q_2$. In the multi-Regge kinematics these replacements 
imply that $I_{4B} = I_{4A}(s_1 \rightarrow s_2,
\vec q_1 \longleftrightarrow - \vec q_2)$.

\section{Appendix}

In this Appendix we derive the relations~(\ref{21}) and~(\ref{22}). Let us 
consider first Eq.~(\ref{21}). From the definition of ${\cal F}_5^\mu$ 
given in the first of Eqs.~(\ref{17}) and from the expression for $I_5^\mu$
given in Eq.~(\ref{I5^m}), we obtain
\begin{equation}
{\cal F}_5^\mu = \frac{s}{\pi^{2+\epsilon}\Gamma(1-\epsilon)}
\biggl[q_1^\mu I_5 - p_A^\mu \frac{I_{4A}}{s} + p_B^\mu \frac{I_{4B}}{s} - I_5^\mu
\biggr]+\frac{1}{2}\ln\left(\frac{s(-s_1)(-s_2)}{(-s)s_1s_2}\right) 
{\cal I}_3^\mu\;.
\label{B1}
\end{equation}
If we contract both sides of the above equation with $2 k_\mu$, we get
\[
2 k_\mu {\cal F}_5^\mu = \frac{s}{\pi^{2+\epsilon}\Gamma(1-\epsilon)}
\biggl[(t_1-t_2) I_5 - \frac{s_1}{s} I_{4A} + \frac{s_2}{s} I_{4B} 
+ I_4^{(1)} - I_4^{(2)}\biggr]
\]
\begin{equation}
+\frac{1}{2}\ln\left(\frac{s(-s_1)(-s_2)}{(-s)s_1s_2}\right)
\left[\frac{\Gamma^2(\epsilon)}{\Gamma(2\epsilon)} 
\biggl((\vec q_2^{\:2})^{\epsilon-1} - (\vec q_1^{\:2})^{\epsilon-1}\biggr)
+(t_1-t_2) {\cal I}_3\right]\;,
\label{Bf-}
\end{equation}
where we have used the relations
\begin{equation}
k_\mu I_5^\mu = \frac{I_4^{(2)} - I_4^{(1)}}{2}\;,
\;\;\;\;\;\;
k_\mu {\cal I}_3^\mu = \frac{1}{2}\frac{\Gamma^2(\epsilon)}{\Gamma(2\epsilon)} 
\biggl((\vec q_2^{\:2})^{\epsilon-1} - (\vec q_1^{\:2})^{\epsilon-1}\biggr)
+\frac{t_1-t_2}{2}{\cal I}_3\;.
\end{equation}
$I_4^{(2)}$ in Eq.~(\ref{Bf-}) is obtained from $I_4^{(1)}$ by the replacements
$p_A \longleftrightarrow p_B$ and $q_1 \longleftrightarrow -q_2$.
In the multi-Regge kinematics these replacements imply $I_4^{(2)} = 
I_4^{(1)}(t_1 \rightarrow t_2, s_2 \rightarrow s_1)$.
Using Eqs.~(\ref{I4_1_fin})-(\ref{I5})
and the third of Eqs.~(\ref{17}) in the R.H.S. of Eq.~(\ref{Bf-}), it is easy 
to obtain the relation~(\ref{21}).

Let us consider now the relation~(\ref{22}). Starting again from Eq.~(\ref{B1})
and contracting both sides with $2 (q_1+q_2)_\mu$, we get 
\[
2 (q_1+q_2)_\mu {\cal F}_5^\mu = \frac{s}{\pi^{2+\epsilon}\Gamma(1-\epsilon)}
\biggl[(t_1+t_2) I_5 + \frac{s_1}{s} I_{4A} + \frac{s_2}{s} I_{4B} 
- I_4^{(1)} - I_4^{(2)} + 2I_4\biggr]
\]
\begin{equation} 
+\frac{1}{2}\ln\left(\frac{s(-s_1)(-s_2)}{(-s)s_1s_2}\right)
\left[\frac{\Gamma^2(\epsilon)}{\Gamma(2\epsilon)} 
\biggl((\vec q_1^{\:2})^{\epsilon-1} + (\vec q_2^{\:2})^{\epsilon-1}
- 2 (\vec k^2)^{\epsilon-1}\biggr)+(t_1+t_2) {\cal I}_3\right]\;,
\label{Bf+}
\end{equation}
where we have used the relations
\[
(q_1+q_2)_\mu I_5^\mu = \frac{I_4^{(1)} + I_4^{(2)}}{2} - I_4 + t_1 I_5\;,
\]
\begin{equation}
(q_1+q_2)_\mu {\cal I}_3^\mu = 
\frac{1}{2}\frac{\Gamma^2(\epsilon)}{\Gamma(2\epsilon)} 
\biggl((\vec q_1^{\:2})^{\epsilon-1} + (\vec q_2^{\:2})^{\epsilon-1}
- 2 (\vec k^2)^{\epsilon-1}\biggr) +\frac{t_1+t_2}{2}{\cal I}_3\;.
\end{equation}
In the previous expressions a new integral appeared
\begin{equation}
I_4 = \frac{1}{i} \int \, d^D p \, \frac{1}
{[(p+q_1)^2+i\varepsilon][(p+q_2)^2+i\varepsilon]
[(p+p_A)^2+i\varepsilon][(p-p_B)^2+i\varepsilon]} \;.
\end{equation}
It can be evaluated using the procedure illustrated in Appendix~A 
for the integral $I_4^{(1)}$. The result is 
\begin{equation}
I_4 = - \frac{\pi^{2+\epsilon}\Gamma(1-\epsilon)}{s}
\frac{\Gamma^2(\epsilon)}{\Gamma(2\epsilon)}\,(\vec k^2)^{\epsilon-1}\,
\left[\ln\left(\frac{(-s)\vec k^2}{(-s_1)(-s_2)}\right)
+ \psi(\epsilon) -\psi(1-\epsilon)\right]\;.
\label{I4}
\end{equation}
Using Eqs.~(\ref{I4_1_fin})-(\ref{I5}), (\ref{I4}) and 
the third of Eqs.~(\ref{17}) in the R.H.S. of Eq.~(\ref{Bf+}), it is easy to obtain 
the relation~(\ref{22}).

\section{Appendix}

In this Appendix we consider the $\epsilon \rightarrow 0$ limit for
the integrals ${\cal I}_3$, ${\cal I}_{4A}$ and for the combination
${\cal I}_5-{\cal L}_3$. The definitions of these integrals are given 
in~(\ref{15}). The integral ${\cal I}_{4B}$, whose expression for
$\epsilon \rightarrow 0$ is given in Eq.~(\ref{43}), can be obtained from
${\cal I}_{4A}$ by the replacement $\vec q_1 \longleftrightarrow - \vec q_2$.

Let us start from ${\cal I}_3$. This integral was calculated
in the Appendix~II of Ref.~\cite{FL93} to the order $\epsilon^0$. 
We report here its expression
\[
{\cal I}_3 = 
\frac{1}{\vec q_1^{\:2}\vec q_2^{\:2}} \left[\frac{1}{\epsilon}
+\ln\left(\frac{\vec q_1^{\:2} \vec q_2^{\:2}} {\vec k^2}\right)\right]
+\frac{1}{\vec q_1^{\:2}\vec k^2} \left[\frac{1}{\epsilon}
+\ln\left(\frac{\vec q_1^{\:2} \vec k^2} {\vec q_2^{\:2}}\right)\right]
+\frac{1}{\vec q_2^{\:2}\vec k^2} \left[\frac{1}{\epsilon}
+\ln\left(\frac{\vec q_2^{\:2} \vec k^2} {\vec q_1^{\:2}}\right)\right]
\]
\begin{equation}
\simeq \frac{1}{\epsilon}\left[
 \frac{(\vec q_1^{\:2} \vec q_2^{\:2})^{\epsilon-1}}{(\vec k^2)^\epsilon}
+\frac{(\vec q_1^{\:2} \vec k^2)^{\epsilon-1}}{(\vec q_2^{\:2})^\epsilon}
+\frac{(\vec q_2^{\:2} \vec k^2)^{\epsilon-1}}{(\vec q_1^{\:2})^\epsilon}\right]\;,
\label{C1}
\end{equation}
where the last approximated equality holds with accuracy $O(\epsilon)$.

The combination of integrals ${\cal I}_5-{\cal L}_3$ can be expressed
in terms of the $I_5$ and ${\cal I}_3$, according to Eq.~(\ref{I5}):
\begin{equation}
{\cal I}_5-{\cal L}_3 = -\frac{s}{\pi^{2+\epsilon}\,\Gamma(1-\epsilon)}\, I_5
+\ln\left(\frac{(-s) \vec k^2}{(-s_1) (-s_2)} \right)\, {\cal I}_3 \;.
\label{C2}
\end{equation}
The integral $I_5$ was calculated in the Appendix~III of Ref.~\cite{FL93} 
to the order $\epsilon^0$:
\begin{eqnarray}
\label{C3}
&-&\frac{s}{\pi^{2+\epsilon}\, \Gamma(1-\epsilon)}\,I_5 =  \nonumber \\
&-&\frac{1}{\vec q_1^{\:2} \vec q_2^{\:2}}\left[\frac{1}{\epsilon^2}
+\frac{1}{\epsilon}
\ln\left(\frac{(-s)\vec q_1^{\:2}\vec q_2^{\:2}}{(-s_1)(-s_2)}\right)
+\frac{1}{2}
\ln^2\left(\frac{(-s)\vec q_1^{\:2}\vec q_2^{\:2}}{(-s_1)(-s_2)}\right)
+\frac{\pi^2}{3}\right] \\
&+&\frac{1}{\vec q_1^{\:2} \vec k^2}\left[\frac{1}{\epsilon^2}
+\frac{1}{\epsilon}
\ln\left(\frac{(-s_1)(-s_2)\vec q_1^{\:2}}{(-s)\vec q_2^{\:2}}\right)
+\frac{1}{2}
\ln^2\left(\frac{(-s_1)(-s_2)\vec q_1^{\:2}}{(-s)\vec q_2^{\:2}}\right)
-2 L\left(1-\frac{\vec q_1^{\:2}}{\vec q_2^{\:2}}\right)\right] \nonumber \\
&+&\frac{1}{\vec q_2^{\:2} \vec k^2}\left[\frac{1}{\epsilon^2}
+\frac{1}{\epsilon}
\ln\left(\frac{(-s_1)(-s_2)\vec q_2^{\:2}}{(-s)\vec q_1^{\:2}}\right)
+\frac{1}{2}
\ln^2\left(\frac{(-s_1)(-s_2)\vec q_2^{\:2}}{(-s)\vec q_1^{\:2}}\right)
-2 L\left(1-\frac{\vec q_2^{\:2}}{\vec q_1^{\:2}}\right)\right]\;, \nonumber
\end{eqnarray}
where 
\begin{equation}
L(x)=\int_0^x\frac{dt}{t}\ln (1-t)\;.
\label{C4}
\end{equation}
Using in~(\ref{C2}) the expression for ${\cal I}_3$ given in the first line of 
Eq.~(\ref{C1}), we get
\begin{eqnarray}
{\cal I}_5-{\cal L}_3 = 
&-&\frac{1}{\vec q_1^{\:2} \vec q_2^{\:2}}\left[\frac{1}{\epsilon^2}
+\frac{1}{\epsilon}
\ln\left(\frac{\vec q_1^{\:2}\vec q_2^{\:2}}{\vec k^2}\right)
+\frac{1}{2}
\ln^2\left(\frac{\vec q_1^{\:2}\vec q_2^{\:2}}{\vec k^2}\right)
-\frac{\pi^2}{6}\right] \nonumber \\
&+&\frac{1}{\vec q_1^{\:2} \vec k^2}\left[\frac{1}{\epsilon^2}
+\frac{1}{\epsilon}
\ln\left(\frac{\vec q_1^{\:2}\vec k^2}{\vec q_2^{\:2}}\right)
+\frac{1}{2}
\ln^2\left(\frac{\vec q_1^{\:2}\vec k^2}{\vec q_2^{\:2}}\right)-\frac{\pi^2}{2}
-2 L\left(1-\frac{\vec q_1^{\:2}}{\vec q_2^{\:2}}\right)\right] \nonumber \\
&+&\frac{1}{\vec q_2^{\:2} \vec k^2}\left[\frac{1}{\epsilon^2}
+\frac{1}{\epsilon}
\ln\left(\frac{\vec q_2^{\:2}\vec k^2}{\vec q_1^{\:2}}\right)
+\frac{1}{2}
\ln^2\left(\frac{\vec q_2^{\:2}\vec k^2}{\vec q_1^{\:2}}\right)-\frac{\pi^2}{2} 
-2 L\left(1-\frac{\vec q_2^{\:2}}{\vec q_1^{\:2}}\right)\right] \nonumber
\end{eqnarray}
\[
\simeq
-\frac{1}{\vec q_1^{\:2} \vec q_2^{\:2}}\left[\frac{1}{\epsilon^2}
\left(\frac{\vec q_1^{\:2}\vec q_2^{\:2}}{\vec k^2}\right)^\epsilon
-\frac{\pi^2}{6}\right]
+\frac{1}{\vec q_1^{\:2} \vec k^2}\left[\frac{1}{\epsilon^2}
\left(\frac{\vec q_1^{\:2}\vec k^2}{\vec q_2^{\:2}}\right)^\epsilon
-\frac{\pi^2}{2} -2 L\left(1-\frac{\vec q_1^{\:2}}{\vec q_2^{\:2}}\right)\right] 
\]
\begin{equation}
+\frac{1}{\vec q_2^{\:2} \vec k^2}\left[\frac{1}{\epsilon^2}
\left(\frac{\vec q_2^{\:2}\vec k^2}{\vec q_1^{\:2}}\right)^\epsilon
-\frac{\pi^2}{2} -2 L\left(1-\frac{\vec q_2^{\:2}}{\vec q_1^{\:2}}\right)\right]\;,
\label{C5}
\end{equation}
where the last approximated equality holds with accuracy $O(\epsilon)$ and we have
used $\ln(-s)=\ln s - i\pi$ and the analogous relations for $s_1$ and $s_2$,
valid in the $s$-channel physical region.

Finally, let us consider ${\cal I}_{4A}$. According to~(\ref{I4A}), it can be 
written in the following way:
\begin{equation}
{\cal I}_{4A} = -\frac{s_1}{\pi^{2+\epsilon}\,\Gamma(1-\epsilon)}\,I_{4A}
-\frac{\Gamma^2(\epsilon)}{\Gamma(2\epsilon)}(\vec q_1^{\:2})^{\epsilon-1}
\left[\ln\left(\frac{(-s_1)}{\vec q_1^{\:2}}\right)+\psi(1-\epsilon)
-2 \psi(\epsilon)+\psi(2\epsilon)\right]\;.
\label{C6}
\end{equation}
The integral $I_{4A}$ was calculated in the Appendix~III of Ref.~\cite{FL93} 
to the order $\epsilon^0$:
\[
-\frac{s_1}{\pi^{2+\epsilon}\,\Gamma(1-\epsilon)}\,I_{4A} = 
\frac{1}{\vec q_1^{\:2}}\left[ \frac{2}{\epsilon^2}
+\frac{2}{\epsilon}\ln\left(\frac{(-s_1) \vec q_1^{\:2}}{\vec q_2^{\:2}}\right)
-\ln^2 \vec q_2^{\:2} + 2 \ln \vec q_1^{\:2} \ln (-s_1) \right.
\]
\begin{equation}
\left. 
-\pi^2 + 2 L\left(1-\frac{\vec q_2^{\:2}}{\vec q_1^{\:2}}\right)\right]\;.
\label{C7}
\end{equation}
Using this expression in the R.H.S. of Eq.~(\ref{C6}) together with 
the expansion to the order $\epsilon^0$ of the second term in the R.H.S.
of the same equation, we get
\[
{\cal I}_{4A} = \frac{1}{\vec q_1^{\:2}}\left[-\frac{1}{\epsilon^2}
+\frac{1}{\epsilon}\ln \vec q_1^{\:2} 
+\frac{1}{2}\ln^2 \vec q_1^{\:2} -\frac{2}{\epsilon}\ln \vec q_2^{\:2} 
-\ln^2 \vec q_2^{\:2} - \frac{\pi^2}{6} 
+ 2 L\left(1-\frac{\vec q_2^{\:2}}{\vec q_1^{\:2}}\right)\right]
\]
\begin{equation}
\simeq
\frac{1}{\vec q_1^{\:2}}\left[\frac{(\vec q_1^{\:2})^\epsilon
-2 (\vec q_2^{\:2})^\epsilon}{\epsilon^2} 
- \frac{\pi^2}{6} + 2 L\left(1-\frac{\vec q_2^{\:2}}{\vec q_1^{\:2}}\right)\right]
\;,
\label{C8}
\end{equation}
where the last approximated equality holds with accuracy $O(\epsilon)$.

\end{document}